\documentclass[
    aps,prd,
    preprint,        
    nofootinbib,     
]{revtex4-2}

\usepackage[T1]{fontenc}   

\usepackage{mathtools}     
\usepackage{amssymb}
\usepackage{bm}
\usepackage{graphicx}
\graphicspath{{article/figures/}}
\usepackage{array}
\usepackage{dcolumn}
\newcolumntype{d}[1]{D{.}{.}{#1}}  
\usepackage{multirow}

\usepackage{tikz}
\usetikzlibrary{calc,positioning}

\usepackage{xcolor}
\definecolor{DarkBlue}{rgb}{0,0,0.4}
\definecolor{DarkGreen}{RGB}{0,100,0}

\usepackage{acro}
\usepackage{orcidlink}    

\usepackage{soul}

\setstcolor{red}
\renewcommand{\st}[1]{}

\DeclareAcronym{ns}{
  short = NS,
  long  = Neutron Star
}
\DeclareAcronym{wd}{
  short = WD,
  long  = white dwarf
}
\DeclareAcronym{bh}{
  short = BH,
  long  = black hole
}
\DeclareAcronym{gr}{
  short = GR,
  long  = general relativity
}
\DeclareAcronym{gw}{
short = GW,
long = gravitational wave
}
\DeclareAcronym{qcd}{
short = QCD,
long = quantum chromodynamics
}
\DeclareAcronym{eos}{
short = EOS,
long = equation of state
}
\DeclareAcronym{pq}{
short = PQ,
long = Peccei--Quinn
}
\DeclareAcronym{sm}{
short = SM,
long = Standard Model
}
\DeclareAcronym{mr}{
short = MR,
long = Mass-Radius
}
\DeclareAcronym{tov}{
short = TOV,
long = Tolman-Oppenheimer-Volkoff
}
\DeclareAcronym{rmf}{
short = RMF,
long = Relativistic Mean Field
}
\DeclareAcronym{cheft}{
short = $\chi$EFT,
long = Chiral Effective Field Theory
}
\DeclareAcronym{chpt}{
short = $\chi$PT,
long = Chiral Perturbation Theory
}
\DeclareAcronym{dm}{
short = DM,
long = Dark Matter
}
\DeclareAcronym{alp}{
short = ALP,
long = Axion-Like-Particle
}
\DeclareAcronym{bsm}{
short = BSM,
long = Beyond the SM
}
\DeclareAcronym{pns}{
short = PNS,
long = Proto Neutron Star
}
\DeclareAcronym{edm}{
short = EDM,
long = Electric Dipole Moment
}
\DeclareAcronym{lqcd}{
short = LQCD,
long = Light--QCD
}
\DeclareAcronym{kg}{
short = KG,
long = Klein--Gordon
}
\DeclareAcronym{ngs}{
short = NGS,
long = New Ground State
}
\DeclareAcronym{mhd}{
short = MHD,
long = Magneto-Hydrodynamics
}
\DeclareAcronym{mft}{
short = MFT,
long = Mean-Field-Theory
}
\DeclareAcronym{qnm}{
short = QNM,
long = Quasi-Normal-Mode
}
\DeclareAcronym{qpo}{
short = QPO,
long = Quasi-Periodic-Oscillation
}
\DeclareAcronym{ms}{
short = MS,
long = Main-Sequence
}
\DeclareAcronym{pnm}{
short = PNM,
long = Pure Neutron Matter
}
\DeclareAcronym{snr}{
short = SNR,
long = Supernova Remnant
}
\DeclareAcronym{durca}{
short = Durca,
long = Direct Urca
}
\DeclareAcronym{murca}{
short = Murca,
long = Modified Urca
}
\DeclareAcronym{cpbf}{
short = CPBF,
long = Cooper pair breaking and formation
}
\DeclareAcronym{lstw}{
short = LSTW,
long = Light-Shinning-Through-Walls
}
\DeclareAcronym{ckm}{
short = CKM,
long = Cabibbo-Kobayashi-Maskawa
}

\usepackage{hyperref}
\hypersetup{
    colorlinks = true,
    linkcolor  = DarkBlue,
    citecolor  = DarkBlue,
    urlcolor   = DarkBlue,
    breaklinks = true,
    pdftitle   = {Nonradial oscillation modes of axion condensed neutron stars
                  in the Cowling approximation},
    pdfauthor  = {A. Gomez-Banon, P. Pnigouras, J. A. Pons},
}

\begin{document}

\title{Nonradial oscillation modes of axion condensed neutron stars
       in the Cowling approximation}

\author{Antonio G\'omez-Ba\~n\'on\,\orcidlink{0009-0007-2546-5372}}
\author{Pantelis Pnigouras\,\orcidlink{0000-0003-1895-9431}}
\author{Jos\'e A. Pons\,\orcidlink{0000-0003-1018-8126}}
\affiliation{Departament de F\'isica, Universitat d'Alacant, 03690 Alicante, Spain}

\date{\today}

\begin{abstract}
In light QCD axion models, finite-density corrections to the axion potential
can source a nonvanishing axion field inside neutron stars. Extending our previous
analysis of radial oscillations, we study the nonradial ($l=2$) oscillation
spectrum of such axion condensed neutron stars in the relativistic Cowling
approximation, keeping the fluid and axion
perturbations fully coupled and dynamical. The condensate modifies the fluid
spectrum in two ways: the altered equilibrium structure raises the
$f$- and $p$-mode frequencies by up to $\sim20\%$, while the dynamical coupling
to the axion field partially compensates this shift, almost entirely for the
$f$-mode. Fluid modes with frequencies above the axion mass
acquire a new damping channel through axion emission. The $f$-mode remains below this threshold for all
configurations supporting a condensate, whereas the $p$-modes
are damped on timescales that can be as short as $\sim10$~ms, shorter than
their gravitational-wave damping times, except for certain axion masses where
the outgoing axion amplitude nearly vanishes. The spectrum also contains a
family of strongly damped axion-dominated quasinormal modes, with imaginary
parts of order kHz, and a sequence of gravitationally bound axion states
analogous to gravitational atom levels. Below the propagation threshold, the bound
$p$-modes undergo avoided crossings with these levels, exchanging their fluid-
and axion-dominated character.
\end{abstract}

\maketitle

\section{Introduction} \label{sec:Introduction}

The QCD axion is a hypothetical pseudo-Nambu--Goldstone boson associated with
the spontaneous breaking of the Peccei--Quinn symmetry, introduced
to solve the strong $CP$ problem of QCD~\cite{peccei1977,peccei1977a,
weinberg1978,wilczek1978}. Remarkably, the same particle is also a natural dark matter
candidate, since a cold axion population can be produced in the early Universe
through the misalignment mechanism~\cite{preskill1983,abbott1983,dine1983}. Together, these features make the QCD axion one of the most compelling
extensions of the Standard Model, and have motivated a broad search programme, combining experiments with astrophysical observations~\cite{irastorza2018, sikivie2021, caputo2024}.
While most astrophysical searches rely on the production and subsequent escape of axions from hot and dense environments, a qualitatively different effect can arise from the response of the axion field itself to dense nuclear matter.

Finite-density corrections modify the axion potential~\cite{hook2018,balkin2020} and can destabilize its vacuum minimum at sufficiently high baryon densities, favouring a nonvanishing axion field. 
\acp{ns}, whose cores reach densities several times nuclear saturation density, provide a natural environment in which this mechanism can operate. The \ac{ns} may then develop a macroscopic axion condensate, with an equilibrium profile extending outside the stellar core over a characteristic Compton wavelength \[ \lambda_a=\frac{\hbar}{m_a c} \simeq 20\,{\rm km} \left(\frac{10^{-11}\,{\rm eV}}{m_a}\right), \] where $m_a$ is the vacuum axion mass. Such condensates can modify the stellar structure and thermal evolution~\cite{hook2018a,diluzio2021,balkin2024,balkin2025,gomez-banon2026,gomez-banon2024,
kumamoto2025}, and are particularly relevant in models in which the axion mass is parametrically suppressed relative to the standard QCD relation.

In our previous work \cite{gomez-banon2026} (hereafter Paper I)
we investigated the radial oscillations of these axion condensed \acp{ns}. Besides the modification of the ordinary fluid-mode frequencies caused by the altered equilibrium configuration, the dynamical axion field was shown to introduce a second family of axion-dominated modes and a new damping channel for the fluid oscillations. Sufficiently far from the star, perturbations of the axion
condensate obey the massive-wave dispersion relation
\begin{equation}\label{eq:disp_rel}
    k=\sqrt{\omega^2 - m_a^2} ,
\end{equation}
in units $\hbar=c=1$. Fluid modes with $\omega>m_a$ can therefore transfer
energy to propagating axion perturbations, whereas below this threshold the
axion perturbation is evanescent and no energy is radiated to infinity through
this channel. The axion mass
thus sets a threshold frequency
\[
    \nu_a\equiv\frac{m_a}{2\pi}\simeq 2.4~\mathrm{kHz}
    \left(\frac{m_a}{10^{-11}~\mathrm{eV}}\right),
\]
which lies in the characteristic kilohertz range of \ac{ns} oscillation
modes for the light axion masses considered here. Although radial modes provide
a clean setting in which this mechanism can be identified, they do not emit
\acp{gw} and are therefore difficult to access observationally.

Nonradial oscillations provide a more promising observational channel. In
particular, modes with multipole number $l\geq2$ couple to gravitational
radiation and form the basis of \ac{gw}
asteroseismology~\cite{andersson1998,kokkotas2001}.
Among them, the fundamental $f$-mode typically lies in the ${\rm kHz}$ frequency range and is particularly sensitive to the global properties of the star. Nonradial oscillations of axion condensed \acp{ns} may thus acquire an additional radiative damping channel, potentially modifying both their oscillation frequencies and damping times and, consequently, the relations between mode properties and stellar parameters used in \ac{gw} asteroseismology.

In this work, we extend the analysis of Paper~I to nonradial oscillations of axion condensed \acp{ns}. We adopt the same equilibrium configurations and axion model developed there, while focusing here on the new ingredients introduced by nonradial perturbations. As a first step, we work within the relativistic Cowling approximation, in which metric perturbations
are neglected while the fluid and axion perturbations remain fully coupled and dynamical. 
\ac{gw} emission is therefore absent by construction, allowing us to isolate the changes in the nonradial spectrum and the damping produced specifically by the axion field.

The remainder of this paper is organized as follows. In Sec.~\ref{sec:theory}, we formulate the nonradial oscillation problem and
summarize the assumptions underlying our analysis. In
Sec.~\ref{sec:results}, we present the resulting mode spectrum and discuss the
effects of the axion condensate. Finally, we summarize our conclusions in
Sec.~\ref{sec:conclusions}.

\section{Nonradial oscillations of axion condensed neutron stars}
\label{sec:theory}

\subsection{Background configurations and assumptions}
\label{subsec:background}

We consider the scenario in which the axion field can condense inside
\acp{ns}, giving rise to a new ground state of nuclear matter~\cite{balkin2024}. This possibility can be realized in models where the vacuum QCD contribution
to the axion potential is suppressed by a small parameter $\epsilon$, which
correspondingly suppresses the axion mass. The reduced vacuum potential allows
finite-density effects from baryonic matter to destabilize the vacuum
configuration and source an axion condensate inside the
star~\cite{hook2018,diluzio2021}.

Denoting the dimensionless axion field by $\theta\equiv a/f_a$, where
$a$ is the axion field and $f_a$ is its decay constant, the vacuum axion potential
and the effective nucleon mass are
\begin{align}
    V(\theta) &= \epsilon m_\pi^2 f_\pi^2\left[1-g(\theta)\right],\\
    m^*_N(\theta) &= m_N-\sigma_N\left[1-g(\theta)\right],
\end{align}
respectively, where
\begin{equation}
    g(\theta)\equiv
    \sqrt{1-\beta\sin^2\left(\frac{\theta}{2}\right)},
    \qquad
    \beta\equiv\frac{4z}{(1+z)^2}\simeq0.88.
\end{equation}
Here, $m_\pi$ and $f_\pi$ are the pion mass and decay constant, respectively,
$z$ is the up-to-down-quark mass ratio, $m_N$ is the nucleon mass, and
$\sigma_N$ is the nucleon sigma term.

The stellar background consists of a static, spherically symmetric
perfect fluid coupled to a radial axion profile, with nuclear matter described
by the BSk26 \ac{eos} \cite{goriely2013}. The axion field modifies the stellar equilibrium both by
reducing the effective nucleon mass and through its contribution to the total
energy density and pressure. The resulting background is determined by
the coupled hydrostatic-equilibrium and axion-field equations. We refer the reader to Paper~I for the explicit form of the equilibrium equations and boundary conditions, and
a detailed discussion of the resulting stellar configurations.

For the present analysis, we take these equilibrium solutions as fixed
backgrounds and focus exclusively on their nonradial perturbations. 
As in Paper~I, we approximate the baryon scalar density $n_s$ with the baryon density $n_b$, i.e., $n_s \simeq n_b$, and set $\epsilon=0.1$. Meanwhile, we vary the axion mass $m_a$ over the range relevant for the oscillation spectrum\footnote{The lower end of this range is set by the condensate itself. For axion
masses below \(\nu_a\simeq2.8\)~kHz, the gradient energy of the field outweighs the
finite-density gain, and no condensate is sourced~\cite{hook2018, balkin2025}; see also Paper~I.}.
The background spacetime metric is written as
\begin{equation}\label{eq:stat_metric}
 ds^2=e^{2\Phi(r)}dt^2-e^{2\lambda(r)}dr^2
      -r^2\left(d\vartheta^2+\sin^2\vartheta\,d\varphi^2\right),
\end{equation}
with the background quantities
$\{p,\varepsilon,n_b,\theta,\Phi,\lambda\}$ depending only on the radial coordinate.

\subsection{Nonradial perturbations in the Cowling approximation}
\label{subsec:perturbations}

We now consider linear, polar nonradial perturbations of the equilibrium
configurations described above.
To simplify the coupled fluid-axion problem, we work in the relativistic
Cowling approximation~\cite{cowling1941,counsell2024}, in which metric
perturbations are neglected. This eliminates \ac{gw} emission as a damping
channel and therefore does not capture the full damping of prominent
\ac{gw} emitters such as the $f$-mode. It does, however, allow us to isolate
the damping associated with axion radiation. For ordinary \ac{ns}
oscillations, the Cowling approximation typically reproduces nonradial mode
frequencies with errors of order $10$--$30\%$~\cite{yoshida1997}, with the
accuracy improving significantly for higher multipoles and overtones.

Owing to the spherical symmetry of the background, the angular dependence of
the perturbations can be expanded in spherical harmonics $Y_{lm}(\vartheta,\varphi)$. Assuming a harmonic time dependence $e^{-i\omega t}$, where $\omega$ is the mode frequency, the scalar perturbations of the pressure $p$ and the dimensionless axion field $\theta$ are decomposed as
\begin{equation}
    \delta p(t,r,\vartheta,\varphi)
    = \left(\frac{r}{R}\right)^l\,\delta p(r)\,Y_{lm}(\vartheta,\varphi)e^{-i\omega t},
    \qquad
    \delta\theta(t,r,\vartheta,\varphi)
    = \left(\frac{r}{R}\right)^l\,\delta\theta(r)\,Y_{lm}(\vartheta,\varphi)e^{-i\omega t},
\end{equation}
and analogously for the energy density perturbations $\delta\varepsilon$ and the baryon density perturbations $\delta n_b$. The factor
$(r/R)^l$, where $R$ is the stellar radius, is extracted explicitly so that the radial functions remain finite
at the origin while retaining the same dimensions for all $l$.

For polar perturbations, the Lagrangian fluid displacement can be decomposed
into radial and tangential components as
\begin{equation}
    \xi^r
    = \left(\frac{r}{R}\right)^l\frac{e^{-\lambda}W(r)}{r}Y_{lm}e^{-i\omega t},
    \qquad
    \xi^A
    = -\left(\frac{r}{R}\right)^l\frac{\mathcal{V}(r)}{r^2}\nabla^A Y_{lm}e^{-i\omega t},
\end{equation}
where $A=\{\vartheta,\varphi\}$, $\nabla^A$ denotes the covariant derivative on
the unit two-sphere, and $W$ and $\mathcal{V}$ describe the radial and
angular fluid displacements, respectively. Since the background is
spherically symmetric, the resulting equations are independent of the azimuthal order $m$.

Linearizing the fluid conservation equations and the axion field equation
around the equilibrium configuration then yields the following coupled
system, where primes denote differentiation with respect to $r$:

\begin{subequations}\label{eq:nonrad_pertsys_main}
\paragraph{Momentum conservation (radial component)}
\begin{equation}\label{eq:nonrad_mom_main}
\begin{split}
    \delta p' + \left(\Phi' + \frac{l}{r}\right)\delta p
    = {}& \omega^2\,e^{2(\lambda-\Phi)}\,\frac{e^{-\lambda}}{r}\,
    (\varepsilon+p)\,W
    - \delta\varepsilon\,\Phi' \\
    &- n_b\,\frac{\partial m^*_N}{\partial\theta}
    \left(\delta\theta' + \frac{l}{r}\,\delta\theta\right)
    - \theta'\!\left(\delta n_b\,\frac{\partial m^*_N}{\partial\theta}
    + n_b\,\frac{\partial^2 m^*_N}{\partial\theta^2}\,\delta\theta\right)
\end{split}
\end{equation}

\paragraph{Baryon conservation}
\begin{equation}\label{eq:nonrad_W_main}
\begin{split}
    \bigl[(\varepsilon+p)\,W\bigr]'
    + (\varepsilon+p)\,W\left(\Phi' + \frac{l+1}{r}\right)
    = -re^{\lambda}\Biggl[ &\frac{\delta p}{c_s^2}
    + \frac{(\varepsilon+p)\,\mathcal{V}\,l(l+1)}{r^2} \Biggr]
\end{split}
\end{equation}
Here, $\mathcal{V}$ satisfies the algebraic equation
\begin{equation}
    (\varepsilon+p)\mathcal{V} = -\frac{e^{2\Phi}}{\omega^2}
    \left[\delta p + n_b\,\frac{\partial m^*_N}{\partial\theta}\,
    \delta\theta\right].
\end{equation}

\paragraph{\ac{kg} equation (interior)}
\begin{equation}\label{eq:nonrad_KG_int_main}
\begin{split}
    e^{-2\lambda}\delta\theta''
    &+ e^{-2\lambda}\left(\Phi' - \lambda' + \frac{2(l+1)}{r}\right)
    \delta\theta' \\
    &+ \left[
        \omega^2 e^{-2\Phi} - m_{\mathrm{eff}}^2
        + \frac{l}{r}e^{-2\lambda}(\Phi' - \lambda')
        + \frac{l(l+1)}{r^2}\left(e^{-2\lambda}-1\right)
    \right]\delta\theta \\
    &= \frac{\delta n_b}{f_a^2}\,
    \frac{\partial m_N^*}{\partial\theta}
\end{split}
\end{equation}
\end{subequations}

where we have defined
\begin{equation}
    m_{\mathrm{eff}}^2
    \equiv \frac{1}{f_a^2}\left(\frac{\partial^2 V}{\partial\theta^2} 
    + n_b\,\frac{\partial^2 m^*_N}{\partial\theta^2}\right).
\end{equation}

\subsection{Boundary conditions and exterior axion field}
\label{subsec:boundary}
Regularity at the origin imposes two independent conditions on the
perturbation variables,
\begin{subequations}\label{eq:origin_reg}
\begin{align}
    \delta p(0)
    &=\frac{e^{-2\Phi(0)}}{l}\omega^2
    \left[\left(\varepsilon+p\right)W\right]_{r=0}
    -n_b(0)\left.\frac{\partial m_N^*}{\partial \theta}\right|_{r=0}\delta\theta(0),
    \\
    \delta\theta'(0)&=0.
\end{align}
\end{subequations}

As in the radial problem studied in Paper~I, the central regularity conditions
are complemented by the vanishing of the Lagrangian pressure perturbation at
the stellar surface,
\begin{equation}
    \Delta p(R)=0,
    \label{eq:DeltapR}
\end{equation}
and by the outgoing wave condition on the axion perturbation at large radius (Appendix~\ref{app:exterior}),
which for the bound modes ($\mathrm{Im}\,k>0$) selects the exponentially decaying solution.

For nonradial perturbations, the exterior \ac{kg} equation differs from its
radial counterpart through the angular term $l(l+1)/r^2$, while metric
perturbations are absent in the Cowling approximation. Defining the physical
exterior axion perturbation as
\begin{equation}
    \delta\theta_{\mathrm{ext}}
    \equiv \left(\frac{r}{R}\right)^l\delta\theta ,
\end{equation}
its radial equation reads
\begin{equation}\label{eq:nonrad_KG_ext_main}
\begin{split}
    \delta\theta_{\mathrm{ext}}''
    &+ \left(\Phi'-\lambda'+\frac{2}{r}\right)
    \delta\theta'_{\mathrm{ext}}
- e^{2\lambda} \left(
    m_{\rm eff}^2
    +\frac{l(l+1)}{r^2}
    -\omega^2e^{-2\Phi}
    \right)\delta\theta_{\mathrm{ext}}=0.
\end{split}
\end{equation}

For a given frequency, the exterior equation and the outgoing boundary condition determine the exterior solution only up to an overall
multiplicative constant. The asymptotic boundary condition can therefore be encoded by the
logarithmic derivative
\begin{equation}\label{eq:Lext_def}
    \mathcal{L}_{\rm ext}(r)\equiv
    \frac{\delta\theta'_{\rm ext}}{\delta\theta_{\rm ext}},
\end{equation}
which is independent of this normalization. Matching the exterior solution
to the interior perturbation at the stellar surface then gives the boundary
condition
\begin{equation}\label{eq:theta_surf}
    \left.\frac{\delta\theta'}{\delta\theta}\right|_{R}
    =\mathcal{L}_{\rm ext}(R)-\frac{l}{R}.
\end{equation}
The numerical computation of $\mathcal{L}_{\rm ext}(R)$ is described in
Appendix~\ref{app:exterior}.

\subsection{Eigenvalue problem and numerical implementation}
\label{subsec:numerics}

The set of Eqs.~\eqref{eq:nonrad_pertsys_main},
\eqref{eq:origin_reg}, \eqref{eq:DeltapR}, \eqref{eq:nonrad_KG_ext_main}, and
\eqref{eq:theta_surf} defines the nonradial eigenvalue problem. 
The interior system~\eqref{eq:nonrad_pertsys_main} consists of two first-order equations for $\delta p$ and $(\varepsilon+p)W$, together with a second-order equation for $\delta\theta$. The problem therefore contains 
four integration constants and the unknown eigenfrequency
$\omega$. Regularity at the
origin provides two conditions, while the vanishing of the Lagrangian
pressure perturbation and the matching to the exterior axion solution provide
two additional conditions at the stellar surface. The remaining freedom
corresponds to the arbitrary overall normalization of the linear
perturbations.

Since the perturbation equations contain formally singular terms at the
origin, the numerical solution is initialized at a small but finite radius.
To improve the accuracy of this initialization, we expand the regular
solution about $r=0$. The required next-order equations lead to
\begin{subequations}\label{eq:nonrad_origin}
\begin{align}
    \delta p'(0) &= 0,
    \\
    \bigl[(\varepsilon+p)\,W\bigr]'(0) &= 0,
    \\
    \delta\theta''(0)
    &= \frac{1}{2l+3}\Biggl\{
        \delta\theta(0)\Bigl[
        -\omega^2 e^{-2\Phi(0)}
        - l\,\Phi''(0)
        + l(l+2)\,\lambda''(0)
        + m_{\mathrm{eff}}^2(0)
        \Bigr]
        \nonumber\\
    &\hphantom{{}={}\frac{1}{2l+3}\Biggl\{}
        + \frac{\delta n_b(0)}{f_a^2}
        \frac{\partial m^*_N}{\partial\theta}(0)
        \Biggr\}.
\end{align}
\end{subequations}
Here, $\delta n_b(0)$ is related to $\delta p(0)$ through
\begin{equation}\label{eq:delp_delnb_approx}
    \delta p(0) = c_s^2\,
(\varepsilon+p)\,\delta n_b(0)/n_b.
\end{equation}

We solve this complex boundary-value problem using the same numerical
strategy developed for the radial oscillations in Paper~I,
appropriately generalized to the nonradial system. In particular, the differential equations and boundary conditions are solved
simultaneously using a collocation method, with the complex wavenumber
\(k\) as unknown and the eigenfrequency given by
\(\omega^2=k^2+m_a^2\).
Starting from an initial estimate of $k$, individual eigenmodes are followed continuously as the axion
mass is varied.

The exterior axion solution is constructed by imposing the outgoing asymptotic condition at large radius $\tilde r_{\infty}$ and integrating inward, first on the
asymptotic vacuum background down to a matching radius $\tilde r_{m}$ and then with the full
exterior \ac{kg} equation~\eqref{eq:nonrad_KG_ext_main} down to the stellar surface, as described in
Appendix~\ref{app:exterior}. We have checked that, at the precision given in Sec.~\ref{sec:results}, 
the fluid-mode frequencies are insensitive to these choices, while the residual $r_m$ dependence of the strongly damped axion modes is quantified separately in Table \ref{tab:axion_modes}.

\section{Results}
\label{sec:results}

\begin{figure}[!htb]
    \centering
        \includegraphics[width=0.4\textwidth]{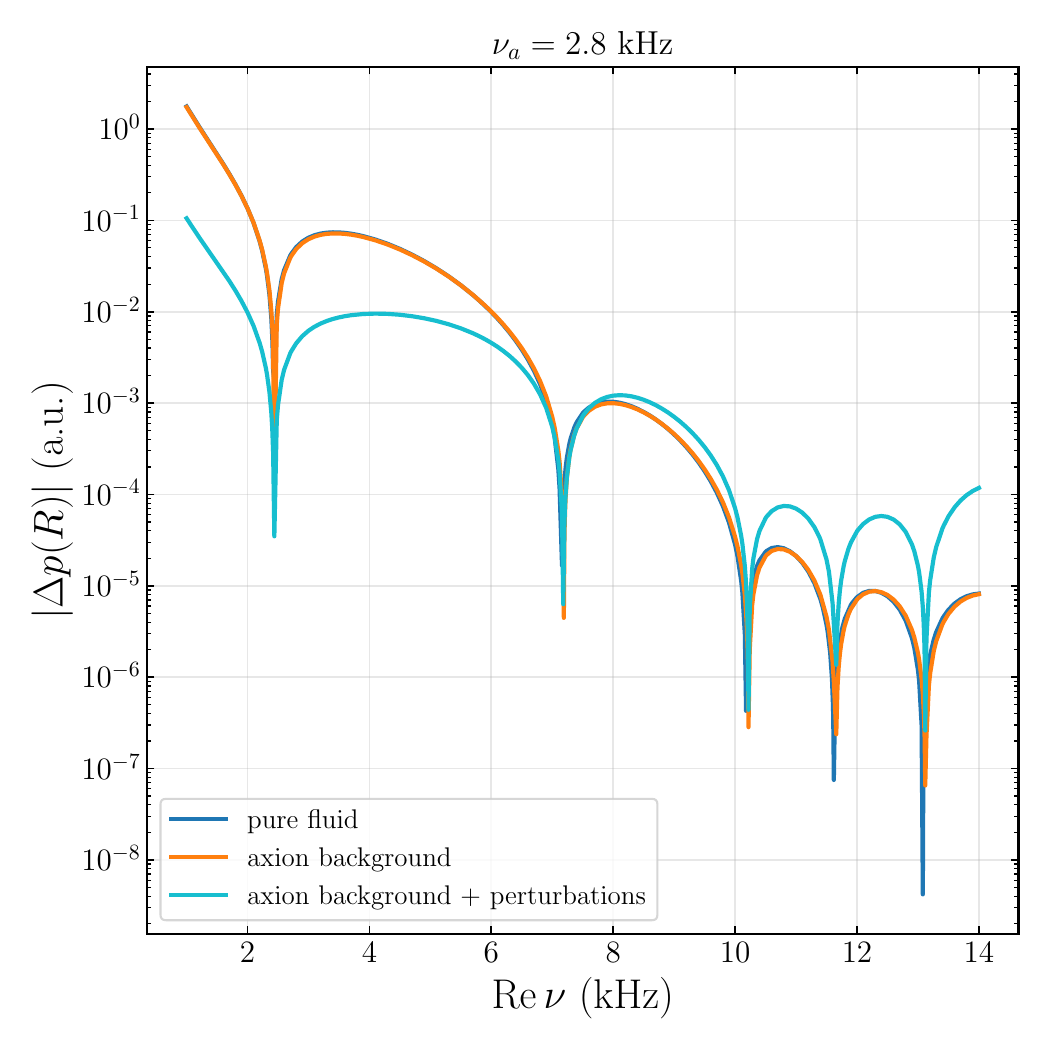} 
        \includegraphics[width=0.4\textwidth]{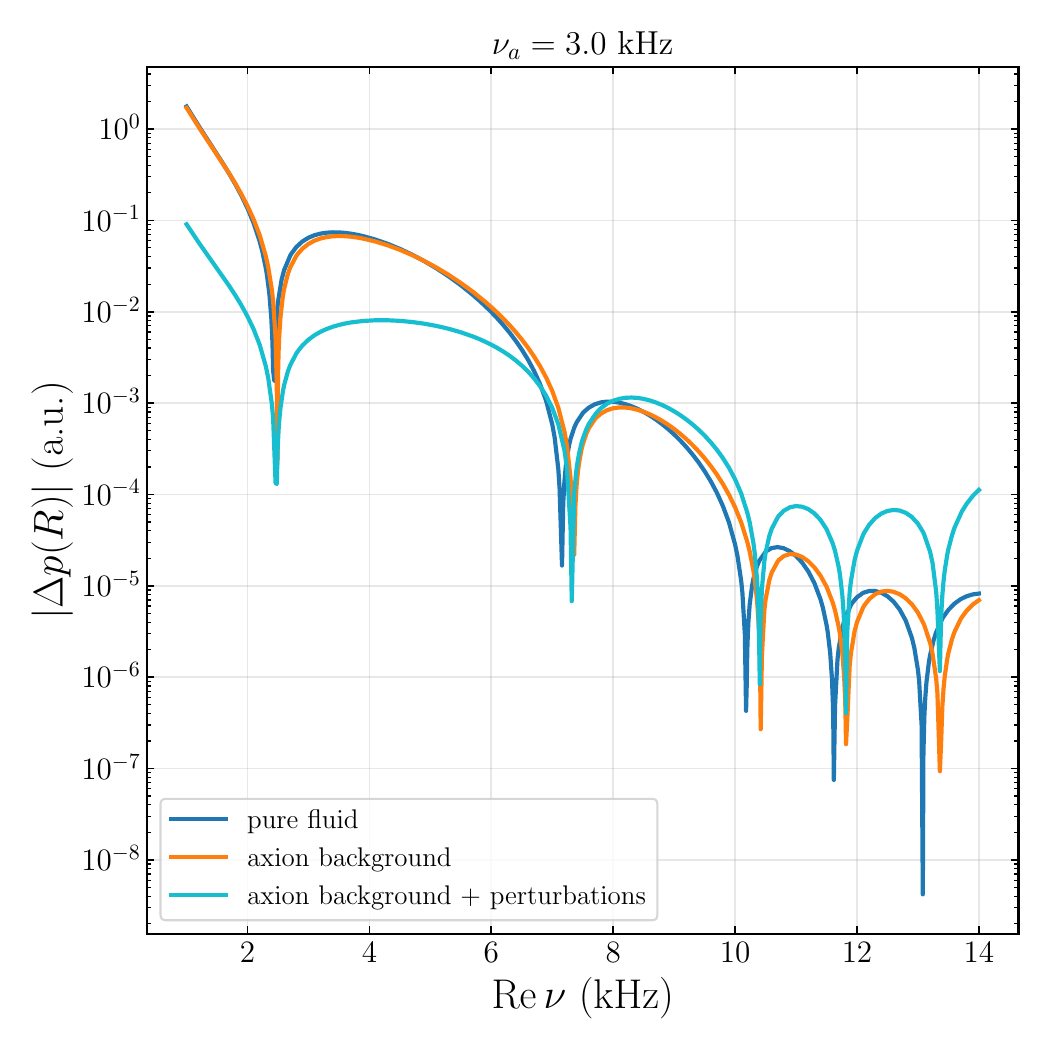} \\
        \includegraphics[width=0.4\textwidth]{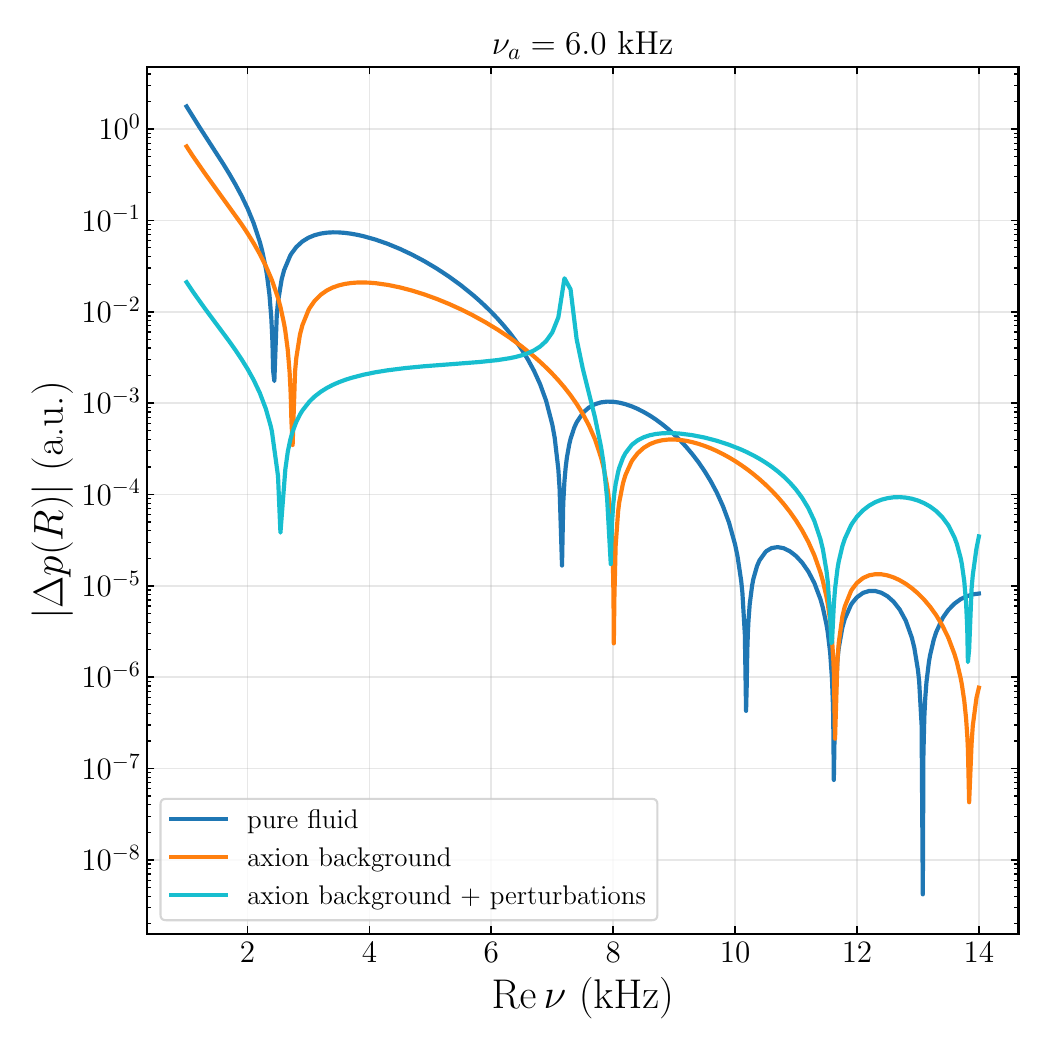} 
        \includegraphics[width=0.4\textwidth]{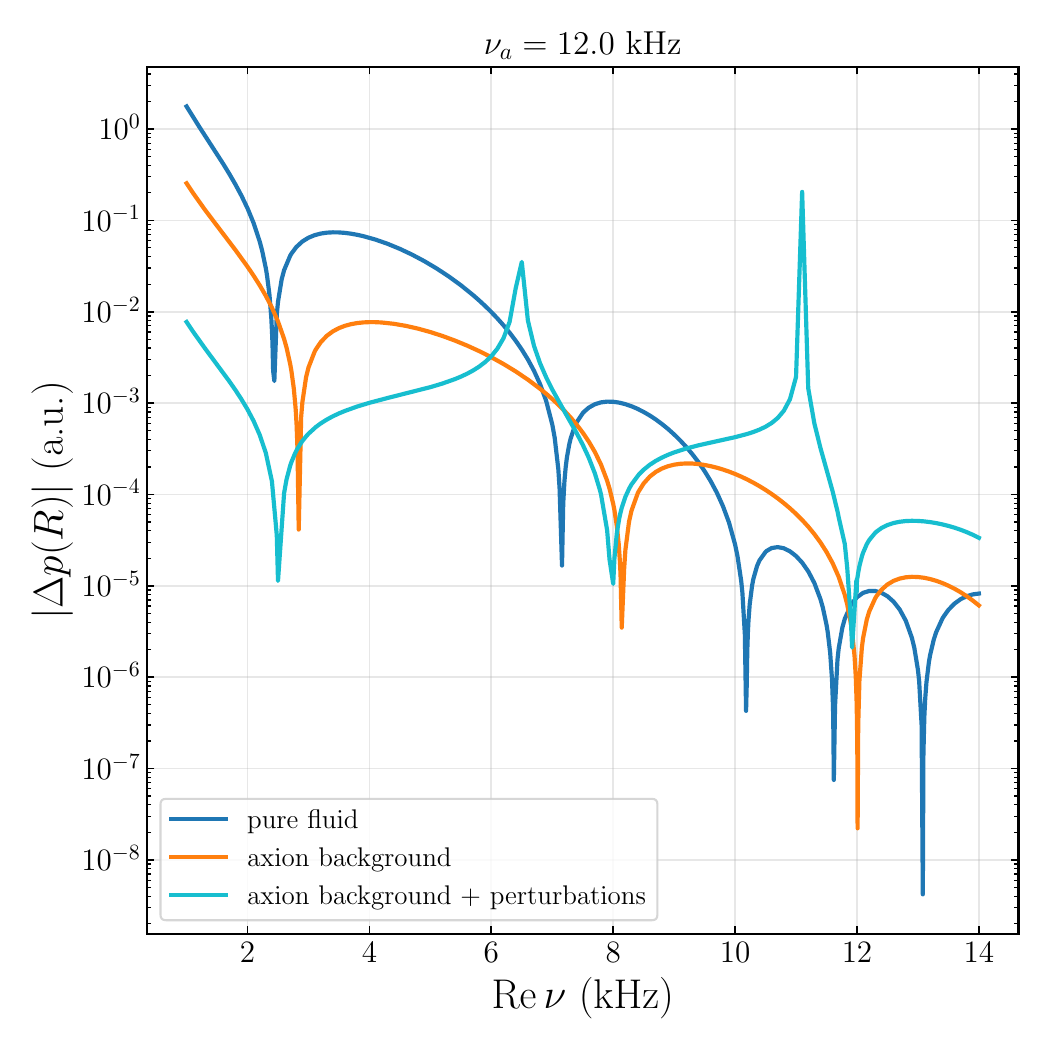} \\
\caption{Surface residual $|\Delta p(R)|$ as a function of the real trial frequency
$\nu=\omega/2\pi$ for $l=2$, $\epsilon=0.1$, and four representative axion masses.
The curves correspond to an ordinary \ac{ns} model without an axion
field (blue), an axion condensed background with axion perturbations
neglected (orange), and the fully coupled fluid--axion system (cyan).
The minima associated with the weakly damped fluid modes provide initial
estimates of their real frequencies, which are subsequently refined by
solving the full complex eigenvalue problem.
}
    \label{fig:fourfigsmodescanp0100eps0.1l2}
\end{figure}

\subsection{Nonradial spectrum of the reference neutron star}

Before analyzing the effects of the axion condensate, we first consider the
nonradial oscillation spectrum of an ordinary neutron star.

The spectrum of nonradial oscillations in relativistic \acp{ns} has been
studied extensively since the pioneering work of Thorne and
Campolattaro~\cite{thorne1967,campolattaro1970}. When the fluid is treated as barotropic, as we assume here, the fluid mode spectrum of the star consists of the fundamental $f$-mode and a
sequence of pressure modes ($p$-modes).%
\footnote{When gradients in the nuclear composition of the \ac{ns} are taken into account, the spectrum is supplemented by a family of buoyancy-driven modes, called $g$-modes.} For typical \ac{ns} models, the
$f$-mode lies in the range $1.5$--$3$~kHz, while the first pressure overtones
appear at several kHz~\cite{andersson1998,kokkotas1999}.

For the reference stellar model considered here, we first compute the
$f$- and $p$-mode frequencies in the absence of the axion field. This
provides both the baseline against which the effects of the axion condensate
will be measured and a validation of our numerical implementation against
standard Cowling calculations.

The weak damping of the fluid modes also provides a simple way of locating
them before solving the full complex eigenvalue problem. We evaluate the
surface residual $|\Delta p(R)|$ by scanning the real frequency axis, while
integrating the perturbation equations for fixed real values of $\omega$.
As illustrated in Fig.~\ref{fig:fourfigsmodescanp0100eps0.1l2}, the fluid modes appear as well-defined
minima of $|\Delta p(R)|$. Their locations provide accurate initial guesses
for the subsequent solution of the complex boundary-value problem described
in the previous section. This procedure is particularly useful for following
individual fluid modes as the axion mass is varied.

Figure~\ref{fig:fourfigsmodescanp0100eps0.1l2} shows these frequency scans for $l=2$ and several
representative axion masses. For comparison, we display three different
calculations: an ordinary \ac{ns} without an axion field, an
axion condensed background in which the dynamical axion perturbation is
neglected, and the fully coupled fluid--axion system. 
The first case identifies the reference $f$- and $p$-mode spectrum, whereas comparison with
the other two cases anticipates the two distinct effects of the axion:
the modification of the equilibrium stellar structure and the dynamical
coupling of the fluid modes to the axion perturbation. These effects will be
quantified separately in the following subsections.

For our reference stellar model, the $f$- and first $p$-mode
frequencies in the absence of the axion field are reported in the first row
of Table~\ref{tab:fluid_modes_axion_backgroundl2}. We obtain
$\nu_f=2.43$~kHz, $\nu_{p_1}=7.15$~kHz, and $\nu_{p_2}=10.18$~kHz.

The agreement of the boundary value problem solver
with previous Cowling calculations~\cite{kunjipurayil2022} is within 0.3\% for the BSk25 \ac{eos}, confirming the accuracy of
the implementation and providing a benchmark for the calculations presented
below.

\begin{table}
    \centering
    \setlength{\tabcolsep}{8pt}
    \begin{tabular}{c c c c}
        \hline\hline
        \(\nu_a~(\mathrm{kHz})\) & \(\nu_f\) & \(\nu_{p_1}\) & \(\nu_{p_2}\) \\
        \hline
        no axion & 2.43 & 7.15 & 10.18 \\
        2.80     & 2.44 & 7.19 & 10.22 \\
        3.00     & 2.49 & 7.35 & 10.42 \\
        4.00     & 2.64 & 7.80 & 11.17 \\
        5.00     & 2.70 & 7.95 & 11.49 \\
        6.00     & 2.73 & 8.01 & 11.64 \\
        8.00     & 2.78 & 8.08 & 11.82 \\
        12.00    & 2.84 & 8.15 & 12.01 \\
        15.00    & 2.87 & 8.19 & 12.11 \\
        \hline\hline
    \end{tabular}
    \caption{Frequencies of the $l=2$ fundamental ($f$) mode and the first two pressure modes ($p_1$ and $p_2$) for the reference \ac{ns} model and for axion condensed configurations with 
    $\epsilon=0.1$ and different $\nu_a$. For the latter, only the modification of the equilibrium background by the axion condensate is included, while dynamical axion perturbations are neglected. All frequencies are given in kHz.}
    \label{tab:fluid_modes_axion_backgroundl2}
\end{table}

\subsection{Axion effects on the fluid modes}
\begin{figure}[!htb]
    \centering
    \includegraphics[width=0.7\linewidth]{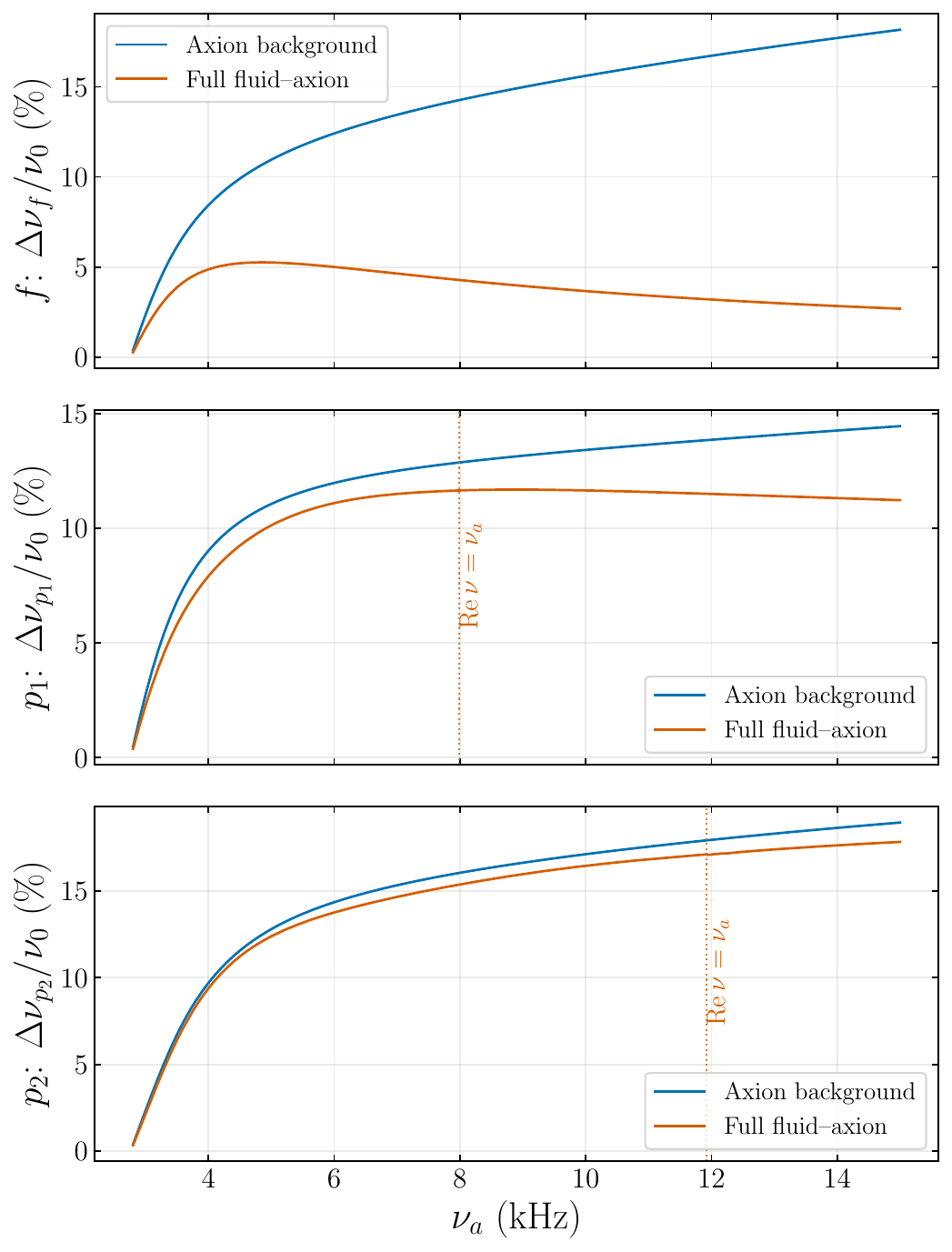}
    \caption{Fractional frequency shifts $\Delta \nu/\nu_{0}$, with
$\Delta \nu = \mathrm{Re}\,\nu - \nu_{0}$, of the $l=2$ $f$, $p_1$ and
$p_2$ modes as a function of $\nu_a$. The reference $\nu_{0}$ is the frequency of the same mode without
axions. Blue: fluid modes on the axion-modified background, with the
axion perturbations switched off. Orange: the fully coupled
fluid--axion system. For the $p$-modes the dotted vertical line marks
the threshold $\mathrm{Re}\,\nu = \nu_a$: for larger $\nu_a$ axion emission is suppressed and we show the corresponding bound state ($\mathrm{Im}\,\nu = 0$) that appears in the spectrum. The $f$-mode lies below $\nu_a$ over the whole
range and is not radiating.}
    \label{fig:fluid_modes_frequencies}
\end{figure}

Having identified the fluid modes and established the reference spectrum, we
now examine systematically how their frequencies are modified by the axion
condensate. 
We first isolate the structural effect by considering perturbations
of the axion-modified stellar backgrounds while neglecting the dynamical axion
perturbation, $\delta\theta$. 
Representative frequencies are listed in Table~\ref{tab:fluid_modes_axion_backgroundl2}. To better illustrate their dependence on the axion mass, in Fig.~\ref{fig:fluid_modes_frequencies} we show the fractional change in the $f$, $p_1$, and $p_2$ frequencies over a more finely sampled range of $\nu_a$.

The background-only results show a systematic increase of the fluid-mode
frequencies with axion mass. 
This trend is common to the $f$-mode
and the first two $p$-modes, although its magnitude is mode dependent. 
Over the range considered here, the frequencies increase from their no-axion
values of $2.43$, $7.15$, and $10.18$~kHz to $2.87$, $8.19$, and
$12.11$~kHz for the $f$, $p_1$, and $p_2$ modes, respectively, at
$\nu_a=15$~kHz. These correspond to relative shifts of approximately $18\%$,
$15\%$, and $19\%$. 
The increase becomes progressively weaker at larger
$\nu_a$, with the frequencies approaching a more slowly varying regime. As in
the radial case studied in Paper~I, these shifts arise entirely from the
modification of the equilibrium stellar structure by the axion condensate,
since $\delta\theta$ is neglected in this calculation.

We next include the dynamical axion perturbation and solve the full coupled
fluid-axion eigenvalue problem. The resulting frequencies are reported in
Table~\ref{tab:fluid_modes_full_axion_scanl2} and compared with the
background-only results in Fig.~\ref{fig:fluid_modes_frequencies},
which shows the fractional frequency shifts. 
The dynamical coupling
partially compensates the upward frequency shift produced by the modified background. 
This effect is particularly pronounced for the $f$-mode: for
example, at $\nu_a=15$~kHz its frequency changes from $2.43$~kHz in the
reference model to $2.87$~kHz when only the axion-modified background is
included, but decreases to $2.5$~kHz in the fully coupled calculation. The corresponding changes for $p_1$ and $p_2$ are
$7.15\rightarrow8.19\rightarrow7.96$~kHz and
$10.18\rightarrow12.11\rightarrow12.00$~kHz, respectively. 
Thus, the partial cancellation between the structural and dynamical effects already
found for radial oscillations in Paper~I persists for the nonradial fluid
modes, although its relative importance differs considerably among the
modes: the cancellation is large for the $f$-mode and comparatively small for the $p$-modes.

The inclusion of $\delta\theta$ also modifies the dependence of the mode
frequencies on $\nu_a$. Whereas the background-only frequencies increase
monotonically throughout the range considered, the fully coupled $f$-mode
initially increases, reaches values around $2.56$~kHz for
$\nu_a\simeq5$--$6$~kHz, and subsequently decreases. 
The effect is less pronounced for the pressure modes, whose
frequencies remain closer to the background-only values. 
This different behavior shows that the dynamical response of the axion field cannot in general be represented as a simple additional frequency shift. The origin of these features will be examined below together with the axion-dominated spectrum.

\subsection{Axion-induced damping of the fluid modes}\label{sec:fluid_damp}

\begin{figure}[!htb]
    \centering
    \includegraphics[width=0.75\linewidth]
    {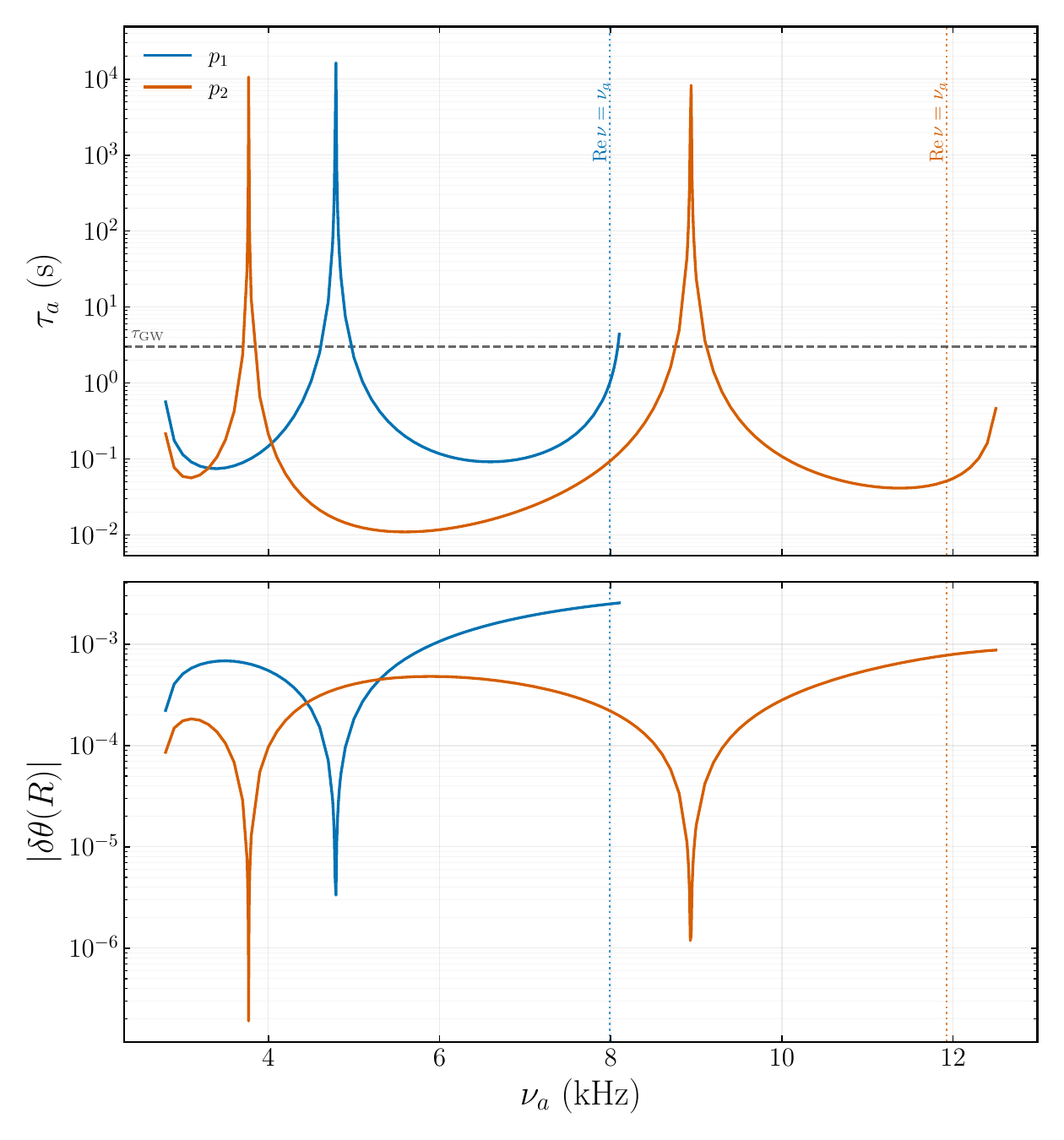}
    \caption{Upper panel: axion-induced damping times $\tau_a$ of the $l=2$ $p_1$ and
$p_2$ modes as a function of $\nu_a$, for $\epsilon=0.1$.
The damping time is defined as
$\tau_a=1/|\mathrm{Im}\,\omega|=1/(2\pi|\mathrm{Im}\,\nu|)$. The horizontal dashed line indicates typical $p$-mode damping time from \ac{gw} emission, $\tau_{\rm GW}$~\cite{kunjipurayil2022}. Lower panel: modulus of the axion perturbation at the surface of the star,
$|\delta\theta(R)|$. The three sharp maxima of $\tau_a$ coincide with the
axion masses at which $\delta\theta(R)$ nearly vanishes, 
indicating a strong suppression of the outgoing axion radiation.
}
\label{fig:axion_damping_times}
\end{figure}

A second consequence of including the dynamical axion field is that the
fluid-mode eigenfrequencies can acquire an imaginary part. As established for
radial oscillations in Paper~I, this occurs when the mode frequency lies above
the axion propagation threshold, $\mathrm{Re}\,\nu>\nu_a$, so that the
perturbation can propagate to infinity and carry energy away from the star.\footnote{
At the threshold, $k$ and $\mathrm{Im}\,\omega$ vanish simultaneously,
and a weakly damped mode admits two continuations below threshold.
The first has $\mathrm{Im}\,k>0$ and an exponentially decaying exterior,
corresponding to the bound states listed in
Table~\ref{tab:fluid_modes_full_axion_scanl2}.
The second extends the damped branch beyond
$\mathrm{Re}\,\nu=\nu_a$ as an anti-bound solution with
$\mathrm{Im}\,k<0$ and an exponentially growing exterior.} The same threshold behavior is apparent in
Table~\ref{tab:fluid_modes_full_axion_scanl2}. The $f$-mode remains below
threshold for all the axion masses considered here and consequently has a
vanishing imaginary part, whereas the $p_1$ and $p_2$ modes acquire finite
damping whenever their frequencies exceed $\nu_a$. 
Once $\nu_a$ becomes larger
than the corresponding mode frequency, the exterior axion perturbation
becomes evanescent and this damping channel closes. 

The propagation threshold determines whether axion emission is kinematically
allowed, but does not by itself determine the magnitude of the damping.
For a weakly damped fluid mode, the radiative losses
can be understood in terms of the asymptotic axion wavenumber $k$ defined in Eq.~\eqref{eq:disp_rel} and the amplitude $A_{\rm out}$ of the outgoing axion wave. To leading
order (see Appendix~\ref{app:power}), the radiated power scales as
\begin{equation}\label{eq:La}
    L_a \propto \mathrm{Re}\,\omega\, \mathrm{Re}\,k|A_{\rm out}|^2,
\end{equation}
while energy balance gives
\begin{equation}\label{eq:La_balance}
    |\mathrm{Im}\,\omega|\simeq\frac{L_a}{2E_{\rm mode}},
\end{equation}
where $E_{\rm mode}$ is the energy stored in the oscillation.
The damping can therefore be suppressed in two physically distinct ways.
Close to the propagation threshold, $\mathrm{Re}\,k\rightarrow0$ and the
radiative losses vanish kinematically as the group velocity of the
outgoing axion tends to zero. In Fig.~\ref{fig:axion_damping_times}, this corresponds to the
approach to the vertical lines from the radiative side.
Away from the threshold, however, the
damping can also become small if the coupled eigenmode excites only a
very small outgoing axion amplitude, $A_{\rm out}\simeq0$. This second possibility is directly illustrated by the lower panel of
Fig.~\ref{fig:axion_damping_times}, where we show the modulus of the axion
perturbation at the stellar surface, $|\delta\theta(R)|$. 
This quantity provides a useful proxy
for the efficiency with which the fluid oscillation excites the exterior
axion field. The three narrow maxima of $\tau_a$ coincide
with deep minima of $|\delta\theta(R)|$, indicating that the strong
suppression of the damping at these particular values of $\nu_a$ is
associated with an almost vanishing axion perturbation at the surface.

The axion-induced damping times obtained here can be compared with the
characteristic damping times associated with \ac{gw} emission in
the standard nonradial problem. Typical \ac{gw} damping times for the first
$p$-modes are of order seconds, although their precise values depend on the
stellar model and on the mode considered~\cite{andersson1998,kokkotas1999}.
The comparison shows that axion
radiative losses can, for a range of values of $\nu_a$, provide a damping channel 
faster than \ac{gw} emission, with the shortest
axion-induced damping times found here reaching down to about ten milliseconds.

It should be stressed that \ac{gw} damping is absent from the
present calculation as a consequence of the Cowling approximation. 
The comparison with $\tau_{\rm GW}$ in Fig.~\ref{fig:axion_damping_times}
is therefore intended  only to indicate the relative importance of the two dissipation 
channels and does not represent the total damping time of the mode.
In a treatment including metric perturbations, the
two radiative channels would operate simultaneously.
If both mechanisms remain weak and can be
treated approximately independently, their damping rates would add, i.e.,
\begin{equation}
    \frac{1}{\tau_{\rm tot}}
    \simeq
    \frac{1}{\tau_{\rm GW}}
    +
    \frac{1}{\tau_a}.
\end{equation}
A calculation beyond the Cowling approximation would be required to test
this approximation and to determine the total damping consistently.

\begin{table}
    \centering
    \begin{tabular}{c c c c c c c}
        \hline\hline
        \multirow{2}{*}{\(\nu_a~(\mathrm{kHz})\)}
        & \multicolumn{2}{c}{\(f\)}
        & \multicolumn{2}{c}{\(p_1\)}
        & \multicolumn{2}{c}{\(p_2\)} \\
        \cline{2-7}
        & \(\mathrm{Re}\,\nu~(\mathrm{kHz})\)
        & \(\mathrm{Im}\,\nu~(\mathrm{Hz})\)
        & \(\mathrm{Re}\,\nu~(\mathrm{kHz})\)
        & \(\mathrm{Im}\,\nu~(\mathrm{Hz})\)
        & \(\mathrm{Re}\,\nu~(\mathrm{kHz})\)
        & \(\mathrm{Im}\,\nu~(\mathrm{Hz})\) \\
        \hline
        2.80  & 2.44 & 0 & 7.18 & \(-0.28\)  & 10.22 & \(-0.74\) \\
        3.00  & 2.47 & 0 & 7.32 & \(-1.4\)   & 10.40 & \(-2.7\) \\
        4.00  & 2.55 & 0 & 7.72 & \(-1.1\)   & 11.14 & \(-0.75\) \\
        5.00  & 2.56 & 0 & 7.88 & \(-0.074\) & 11.44 & \(-12\) \\
        6.00  & 2.55 & 0 & 7.95 & \(-1.4\)   & 11.58 & \(-13\) \\
        7.00  & 2.54 & 0 & 7.98 & \(-1.6\)   & 11.68 & \(-7.1\) \\
        8.00  & 2.54 & 0 & 7.99 & 0          & 11.75 & \(-1.7\) \\
        10.00 & 2.52 & 0 & 7.99 & 0          & 11.86 & \(-1.5\) \\
        11.00 & 2.51 & 0 & 7.98 & 0          & 11.90 & \(-3.8\) \\
        12.00 & 2.51 & 0 & 7.98 & 0          & 11.91 & 0 \\
        13.00 & 2.50 & 0 & 7.97 & 0          & 11.95 & 0 \\
        15.00 & 2.50 & 0 & 7.96 & 0          & 12.00 & 0 \\
        \hline\hline
    \end{tabular}
    \caption{Real and imaginary parts of the frequencies of the \(l=2\) fundamental mode \(f\) and the first two pressure modes, \(p_1\) and \(p_2\), for different \(\nu_a\). The real parts are given in kHz, while the imaginary parts are given in Hz. All exterior solutions are obtained with the phase--amplitude method of
Appendix~\ref{app:exterior}. Direct inward integration of
Eq.~\eqref{eq:nonrad_KG_ext_main} from \(\tilde r_\infty\) reproduces all
entries to better than \(2\times10^{-4}\) in \(\mathrm{Re}\,\nu\) and, for the
damped modes, to within \(1.5\%\) in \(\mathrm{Im}\,\nu\).}
    \label{tab:fluid_modes_full_axion_scanl2}
\end{table}

\subsection{Nonradial axion modes}
\label{sec:axion_modes}

The coupled system also admits a distinct family of axion-dominated
\acp{qnm}, already present in the \ac{kg} spectrum on the fixed fluid-axion background. These modes persist when the dynamical coupling to the fluid is included and are characterized by comparatively large imaginary parts. In this respect, they are qualitatively
reminiscent of the spacetime $w$-modes of relativistic stars, which are also weakly coupled to the fluid and have no Newtonian analog
\cite{kokkotas1992}.

\begin{figure}[!htb]
\centering
\includegraphics[width=0.8\linewidth]{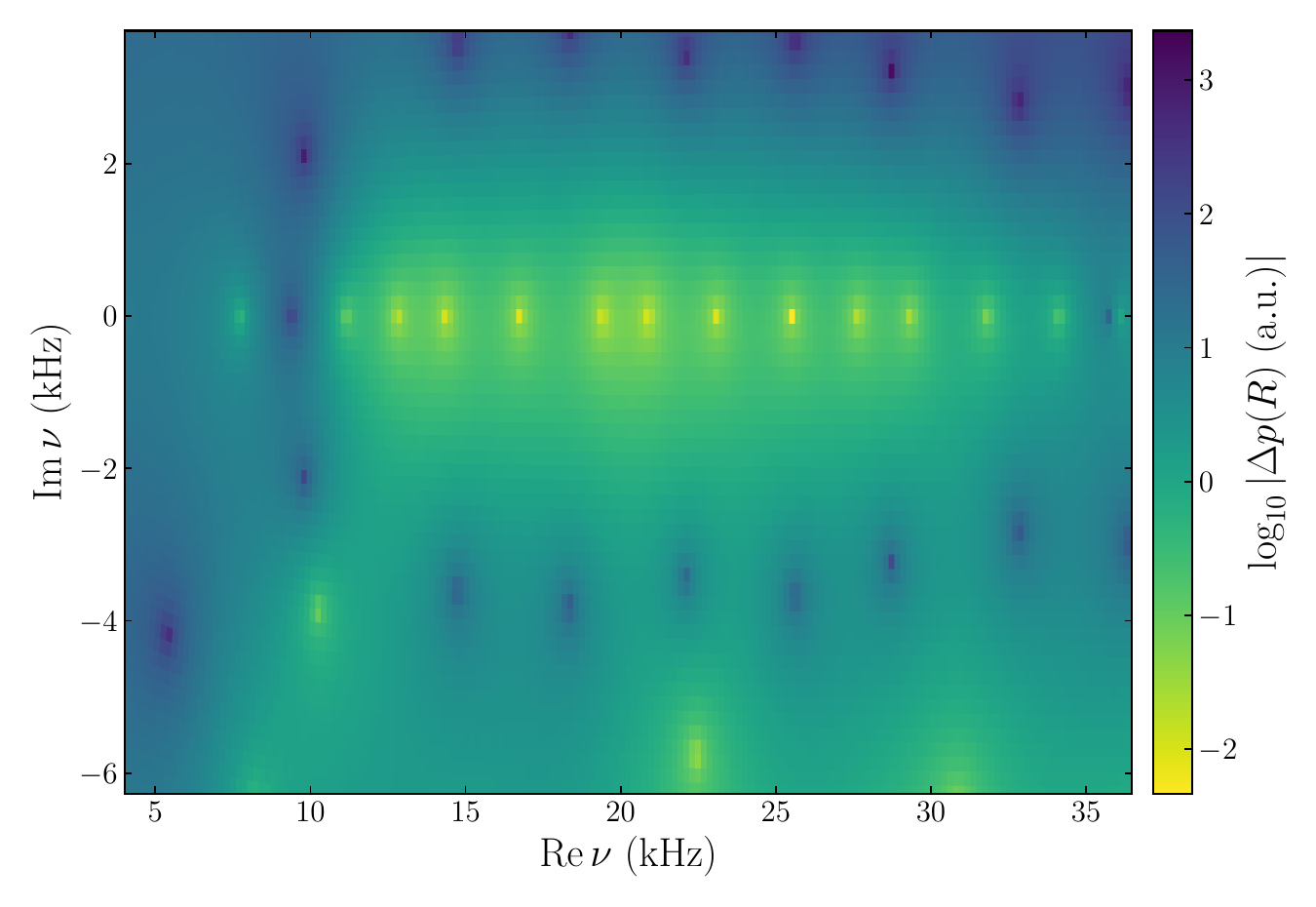}
\caption{Complex-frequency landscape for $\nu_a=4.0~\mathrm{kHz}$,
represented by the minima of $\log_{10}|\Delta p(R)|$. Local minima (yellow in the color scale) identify candidate \acp{qnm} and provide initial guesses for the full boundary-value solver. 
Two qualitatively different families are visible:
the weakly damped fluid-dominated modes, which lie close to the real axis (see Table~\ref{tab:fluid_modes_full_axion_scanl2}), and the strongly damped axion-dominated modes.
In the frequency range shown here, the latter are located at
$\nu\simeq 10.2-3.8\,i~\mathrm{kHz}$ and
$\nu\simeq22.4-5.7\,i~\mathrm{kHz}$,
corresponding to the $a_0$ and $a_1$ modes listed in Table~\ref{tab:axion_modes}. 
The left vertical boundary at $\mathrm{Re}\,\nu=\nu_a$ marks the real frequency threshold.
}
\label{fig:2d_modes}
\end{figure}

Because the axion-dominated modes are strongly damped, the real-frequency
scans used to locate the weakly damped fluid modes are not suitable for
identifying them. Following the strategy adopted for radial oscillations
in Paper~I, we therefore explore the complex-frequency plane and search
for minima of the residuals associated with the boundary conditions.
These two-dimensional scans provide initial estimates of the complex
eigenfrequencies, which are subsequently refined by solving the full
boundary-value problem. An example is shown in
Fig.~\ref{fig:2d_modes}, while representative frequencies are listed in Table~\ref{tab:axion_modes}.

The two axion-dominated modes exhibit a markedly different dependence on $\nu_a$. The $a_0$ mode remains in the range
$\mathrm{Re}\,\nu\simeq 9$--$13~\mathrm{kHz}$ over the models considered, with a non-monotonic variation of its real frequency and a damping rate that decreases substantially at the largest axion masses. 
By contrast, the $a_1$ branch increases steadily in frequency, from
$\mathrm{Re}\,\nu\simeq19~\mathrm{kHz}$ at $\nu_a=2.8~\mathrm{kHz}$ to
more than $50~\mathrm{kHz}$ at $\nu_a=15~\mathrm{kHz}$. Both are much more strongly damped than the fluid modes: their imaginary parts are typically of order kHz rather than Hz. This confirms that they represent a genuinely different sector of the spectrum, dominated by the axion degree of freedom rather than by fluid motion.
\begin{table}
    \centering
    \begin{tabular}{c c c c c c c}
        \hline\hline
        \multirow{2}{*}{\(\nu_a~(\mathrm{kHz})\)}
        & \multicolumn{3}{c}{$a_0$}
        & \multicolumn{3}{c}{$a_1$} \\
        \cline{2-7}
        & \(\mathrm{Re}\,\nu\) & \(\mathrm{Im}\,\nu\) & \(\left|\delta \nu_m\right|\)
        & \(\mathrm{Re}\,\nu\) & \(\mathrm{Im}\,\nu\) & \(\left|\delta \nu_m\right|\) \\
        \hline
        2.80  &  9.920 & $-2.623$ & 0.001 & 18.87 & $-5.49$ & 0.02 \\
        3.00  & 10.15  & $-2.86$  & 0.01  & 19.56 & $-5.61$ & 0.07 \\
        4.00  & 10.2   & $-3.8$   & 0.2   & 22.42 & $-5.74$ & 0.06 \\
        5.00  &  9.1   & $-4.2$   & 0.4   & 25.29 & $-4.72$ & 0.02 \\
        6.00  &  9.0   & $-3.8$   & 0.3   & 27.95 & $-3.61$ & 0.02 \\
        8.00  &  9.3   & $-3.3$   & 0.3   & 33.18 & $-2.39$ & 0.01 \\
        $12.00^{*}$ & 10.9   & $-2.0$   & 0.3   & 44.68 & $-1.87$ & 0.01 \\
        $15.00^{*}$ & 12.5   & $-0.5$   & 0.2   & 53.67 & $-2.13$ & 0.02 \\
        \hline\hline
    \end{tabular}
    \caption{First two axion-dominated modes. The exterior solution is matched to the
    phase-amplitude integration at $r_m=R+1/m_a$. Beyond this radius the
    residual background axion profile is neglected and the field is evolved on
    the asymptotic vacuum background (see Appendix~\ref{app:exterior_pa}).
    We estimate the corresponding systematic uncertainty as
    $\left|\delta \nu_m\right| \equiv \left|\nu(r_m=R+1/m_a)-\nu(r_m=R+0.5/m_a)\right|,$
    i.e. from the shift obtained by halving the matching distance.
    Each frequency is quoted only to the precision supported by this estimate. All frequencies are in kHz. Modes marked with an asterisk have $\mathrm{Re}\,\nu<\nu_a$. For these
strongly damped modes $\mathrm{Re}\,k$ does not vanish at the threshold but
decreases gradually with $\nu_a$, and their radiative flux is correspondingly
suppressed.}

    \label{tab:axion_modes}
\end{table}

In addition to the axion-dominated modes listed in Table~\ref{tab:axion_modes}, 
there is a second family of gravitationally bound axion states
with $\omega \lesssim  m_a$.
These states are confined by the attractive
$1/r$ tail of the gravitational potential of the \ac{ns}
and are undamped within the conservative
perturbation problem considered here, since 
they cannot radiate axions to infinity 
($k^2<0$ and they decay exponentially at large distances).

\begin{figure}[!htb]
    \centering
    \includegraphics[width=0.8\linewidth]{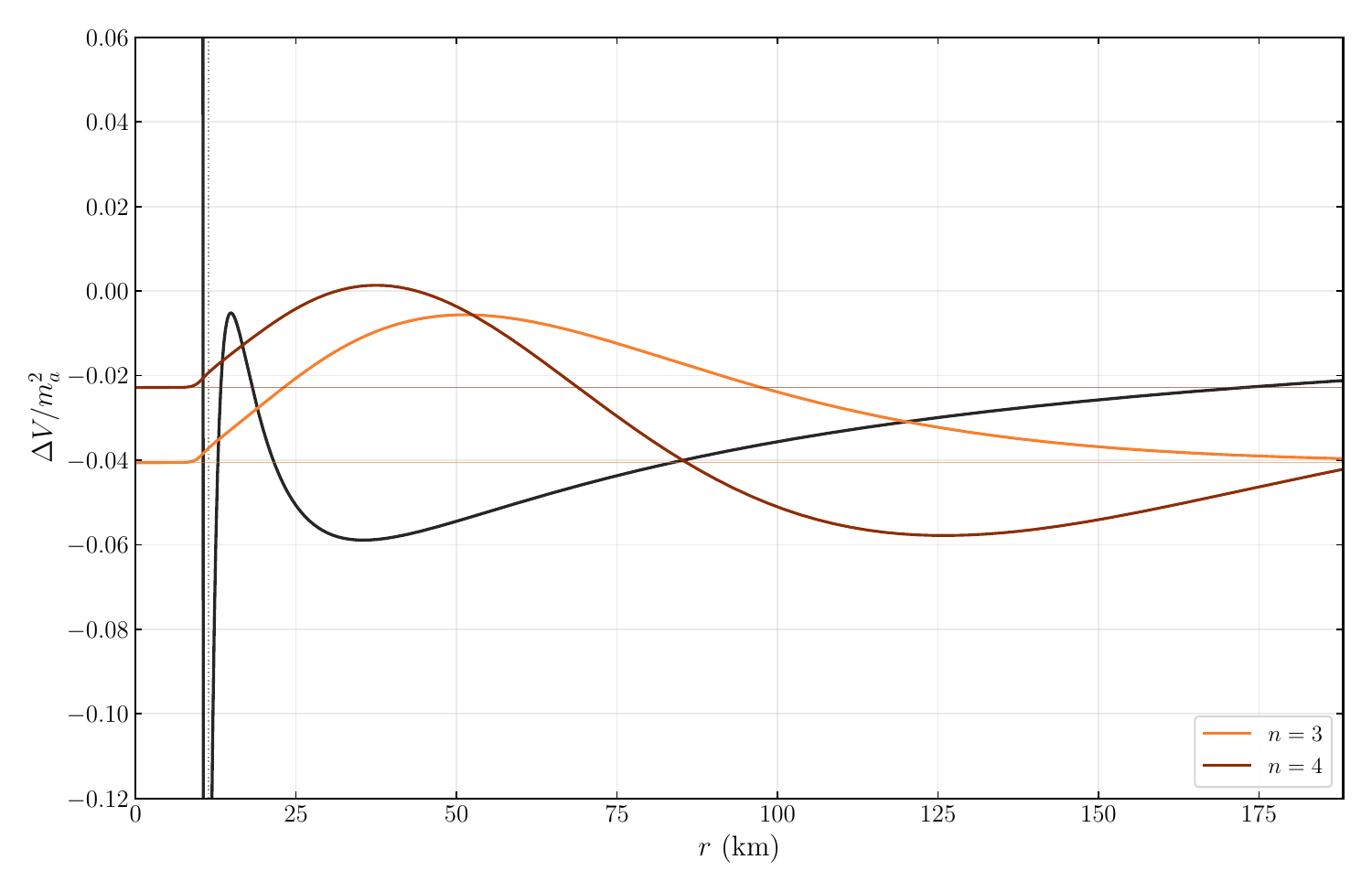}
    \caption{Potential $\Delta V$ of Eq.~\eqref{eq:kg_psi} in units of $m_a^2$
(black), for $\nu_a = 12.0$~kHz, and the exterior axion eigenfunctions
$\psi = r\,\delta\theta_{\rm ext}$ of the $n = 3$ and $n = 4$ gravitational
atom levels. The eigenfunctions have arbitrary amplitudes and are vertically shifted at the height of their $\tilde{k}_n^2$ (thin horizontal lines). The dotted vertical line marks the stellar surface. 
The narrow dip of $\Delta V$ just inside the surface is the region where $m_{\rm eff}^2 < 0$, where the background axion field rapidly changes from $\theta\sim\pi$ to $\theta\sim 0$.
}
\label{fig:axion_levels_eigenfunctions}
\end{figure}

Figure~\ref{fig:axion_levels_eigenfunctions}
shows their radial eigenfunctions superposed on the effective
potential $\Delta V$.
Introducing $\psi=r\,\delta\theta_{\rm ext}$ and the tortoise coordinate $dr^*=e^{\lambda-\Phi}\,dr$,
the exterior perturbed \ac{kg} equation~\eqref{eq:nonrad_KG_ext_main} takes the
Schr\"odinger-like form
\begin{equation}\label{eq:kg_psi}
    \frac{d^2\psi}{dr^{*2}} + \left(k^2 - \Delta V\right)\psi = 0 ,
\end{equation}
where $\Delta V(r\rightarrow\infty)=0$. 
In a Schwarzschild exterior, neglecting corrections from the background axion field, the potential reduces to
\begin{equation}\label{eq:pot_tortoise}
    \Delta V = -\frac{2G M m_a^2}{r}
    + \left(1-\frac{2GM}{r}\right)\left[\frac{l(l+1)}{r^2} + \frac{2GM}{r^3}\right].
\end{equation}
 
At large distances, $r\gg2GM$, the leading terms are
$\Delta V\simeq-2GMm_a^2/r+l(l+1)/r^2$. With $r^*\simeq r$
to leading order, Eq.~\eqref{eq:kg_psi} then becomes the radial
Coulomb problem, yielding the hydrogenic estimate
\begin{equation}\label{eq:bohr}
    \tilde{k}_n^2 \equiv \frac{k_n^2}{m_a^2} = -\frac{\alpha^2}{n^2},
    \qquad \alpha \equiv GMm_a, \qquad n \geq l+1 .
\end{equation}
Equivalently, $\omega_n\simeq m_a[1-\alpha^2/(2n^2)]$. 
These are the ``gravitational atom'' levels familiar from ultralight bosons around compact
objects~\cite{detweiler1980,cannizzaro2026}. 
For the models considered here $\alpha \approx 0.4$--$0.7$
(see Table~\ref{tab:axion_levels}).
Although this coupling is not
asymptotically small, the lowest hydrogenic level for $l=2$ has
$n=3$ and remains weakly bound.
The characteristic radial scale,
$r_n\sim n^2/(\alpha m_a)=n^2GM/\alpha^2$, places these states
predominantly outside the star, as illustrated in
Fig.~\ref{fig:axion_levels_eigenfunctions}.

\begin{table}
  \centering
  \begin{tabular}{ccccccc}
    \hline\hline
    $\nu_a$ (kHz) & $M/M_\odot$ & $\alpha$ & $\nu_{n=3}$ (kHz) & $-\tilde{k}_3^2\,n^2/\alpha^2$
      & $\nu_{n=4}$ (kHz) & $-\tilde{k}_4^2\,n^2/\alpha^2$ \\
    \hline
     8.0 & 1.494 & 0.370 &  7.94 & 1.06 &  7.96 & 1.05 \\
     9.0 & 1.501 & 0.418 &  8.91 & 1.07 &  8.95 & 1.07 \\
    10.0 & 1.507 & 0.466 &  9.87 & 1.10 &  9.93 & 1.09 \\
    11.0 & 1.512 & 0.515 & 10.82 & 1.12 & 10.90 & 1.12 \\
    12.0 & 1.515 & 0.563 & 11.75 & 1.15 & 11.86 & 1.15 \\
    13.0 & 1.519 & 0.611 & 12.68 & 1.19 & 12.82 & 1.19 \\
    14.0 & 1.521 & 0.659 & 13.58 & 1.22 & 13.77 & 1.22 \\
    15.0 & 1.524 & 0.707 & 14.47 & 1.25 & 14.71 & 1.24 \\
    \hline\hline
  \end{tabular}
  \caption{Frequencies of the $n=3$ and $n=4$ gravitationally bound
axion modes with $l=2$, computed with fluid perturbations switched
off. 
The ratios $-\tilde{k}_n^2 n^2/\alpha^2$ measure departures
from the hydrogenic spectrum of Eq.~\eqref{eq:bohr}, for which
they equal unity. 
Including fluid perturbations shifts these
frequencies by less than $3\times10^{-3}\,\mathrm{kHz}$,
below the precision quoted in the table.}
\label{tab:axion_levels}
\end{table}

\subsection{Avoided crossings between gravitational atom levels and $p$-modes}\label{sec:avoid_cross}
The gravitational atom levels are states dominated by the exterior axion field: they exist for every $m_a$, and
are present with the fluid perturbations switched off. They can affect the
fluid modes since their frequencies lie just below $\nu_a$, in the range the
bound $p$-modes sweep through once they cross the radiation threshold. 
Where their uncoupled frequencies would cross, the dynamical coupling produces level repulsion and the two coupled branches exchange their fluid- and axion-dominated character.

\begin{figure}[!htb]
    \centering
    \includegraphics[width=\linewidth]{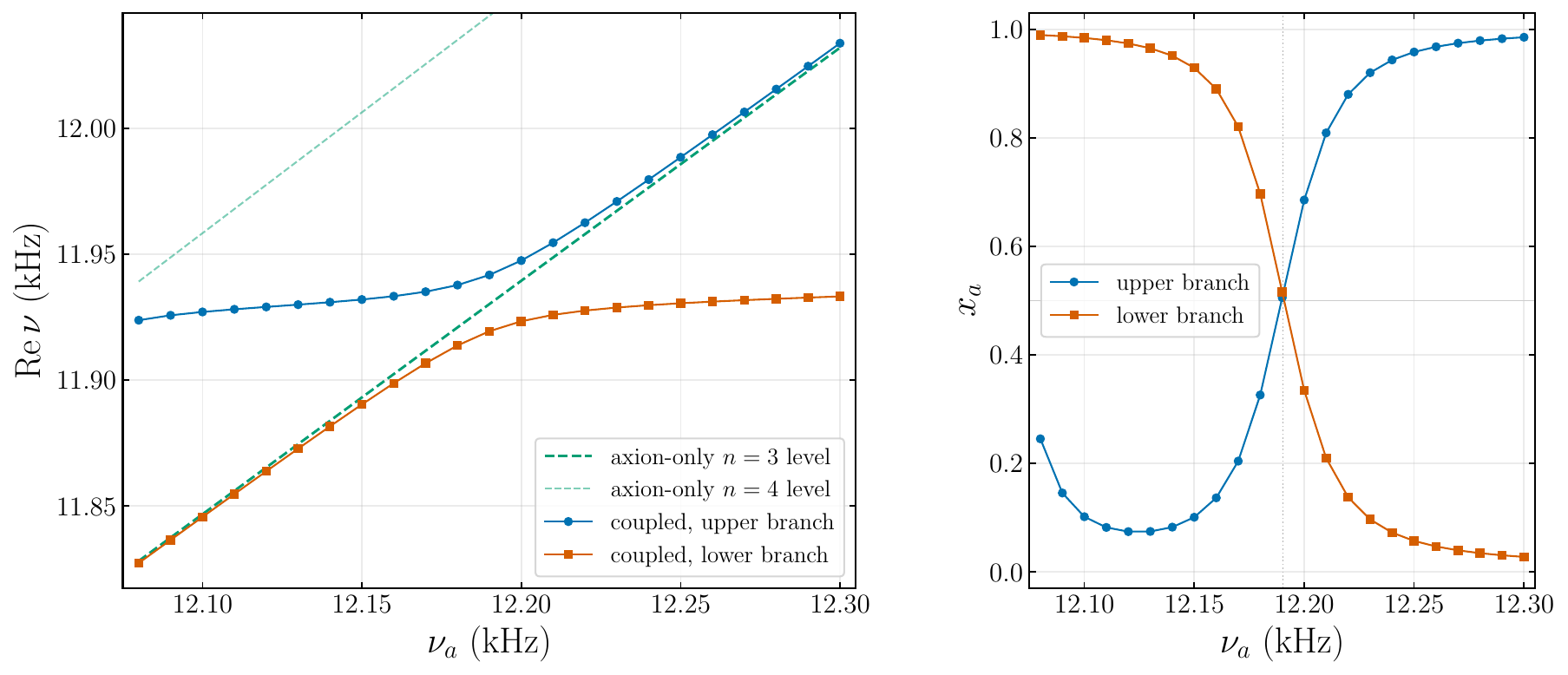}
    \caption{Avoided crossing of the bound $p_2$ mode with the $n = 3$
gravitational atom level. Left: $\mathrm{Re}\,\nu$ of the two coupled branches
(blue, orange) and of the $n = 3$ and $n = 4$ levels of the axion-only
problem (green, dashed), as functions of $\nu_a$. Below $\nu_a \simeq 12.19$~kHz the
upper branch is the fluid-like $p_2$ and the lower one follows the $n = 3$
level, above it they have exchanged character. Right: axion energy fraction
$x_a$ [Eq.~\eqref{eq:xa}], along each branch. The curves cross at
$x_a = 0.5$ at the closest approach. The rise of the upper branch towards low
$\nu_a$ is the beginning of its crossing with the $n = 4$ axion level.}
    \label{fig:p2_n3_crossing}
\end{figure}

In Fig.~\ref{fig:p2_n3_crossing} we plot $\mathrm{Re}\,\nu$ for the $n=3,4$ levels as a function of $\nu_a$, next to the obtained sequences for the $p_2$-mode in the full coupled problem. 
The coupled spectrum contains two branches that undergo an avoided crossing near the intersection of the uncoupled $p_2$-mode and the $n=3$ axion level.
The closest approach occurs at $\nu_a\simeq12.19$ kHz, where the two branches exchange their fluid- and axion-dominated character.
The onset of a similar interaction with the $n=4$ level is already visible at lower $\nu_a$, while resolving the higher levels requires progressively finer sampling.

To quantify the character of a mode across the crossing, we use
a measure of the axion content of the mode,
$x_a=E_a/(E_a+E_{\rm fluid})$ [see Eq.~\eqref{eq:xa}], where $E_{\rm fluid}$ and
$E_a$ are twice the period-averaged kinetic energies of the fluid and of the
axion field, defined in Appendix~\ref{app:power}. 
For a bound mode, their sum
is the total mode energy, so $x_a$ runs from $0$ for a pure fluid mode to $1$ for a pure axion level.
Below $\nu_a \simeq 12.19$~kHz the upper branch is the
fluid-like $p_2$, nearly flat at $11.93$~kHz, and the lower branch follows the
$n = 3$ level, which rises with $\nu_a$. Above it, the roles are exchanged: 
the upper branch follows the axion level and the lower one becomes the $p_2$-mode. The right panel shows $x_a$ along
each branch: it goes from $0.98$ to $0.03$ on the lower branch and from $0.1$
to $0.99$ on the upper one, the two curves crossing at $x_a = 0.5$ at $\nu_a =
12.19$~kHz. The rise of $x_a$ on the
upper branch towards low $\nu_a$ is the start of the next crossing, with the
$n = 4$ level at $\nu_a \simeq 12.06$~kHz. 
Since the gravitational-atom levels accumulate below $\nu_a$, the bound $p_2$ branch is expected, in principle, to encounter an increasingly dense sequence of progressively narrower avoided crossings as the propagation threshold is approached.
We stress that these avoided crossings occur in the bound regime, after the axion-radiation channel has closed, and are therefore distinct from the narrow maxima of the radiative damping time discussed in Sec. \ref{sec:fluid_damp}.

\begin{figure}[!htb]
    \centering
    \includegraphics[width=\linewidth]{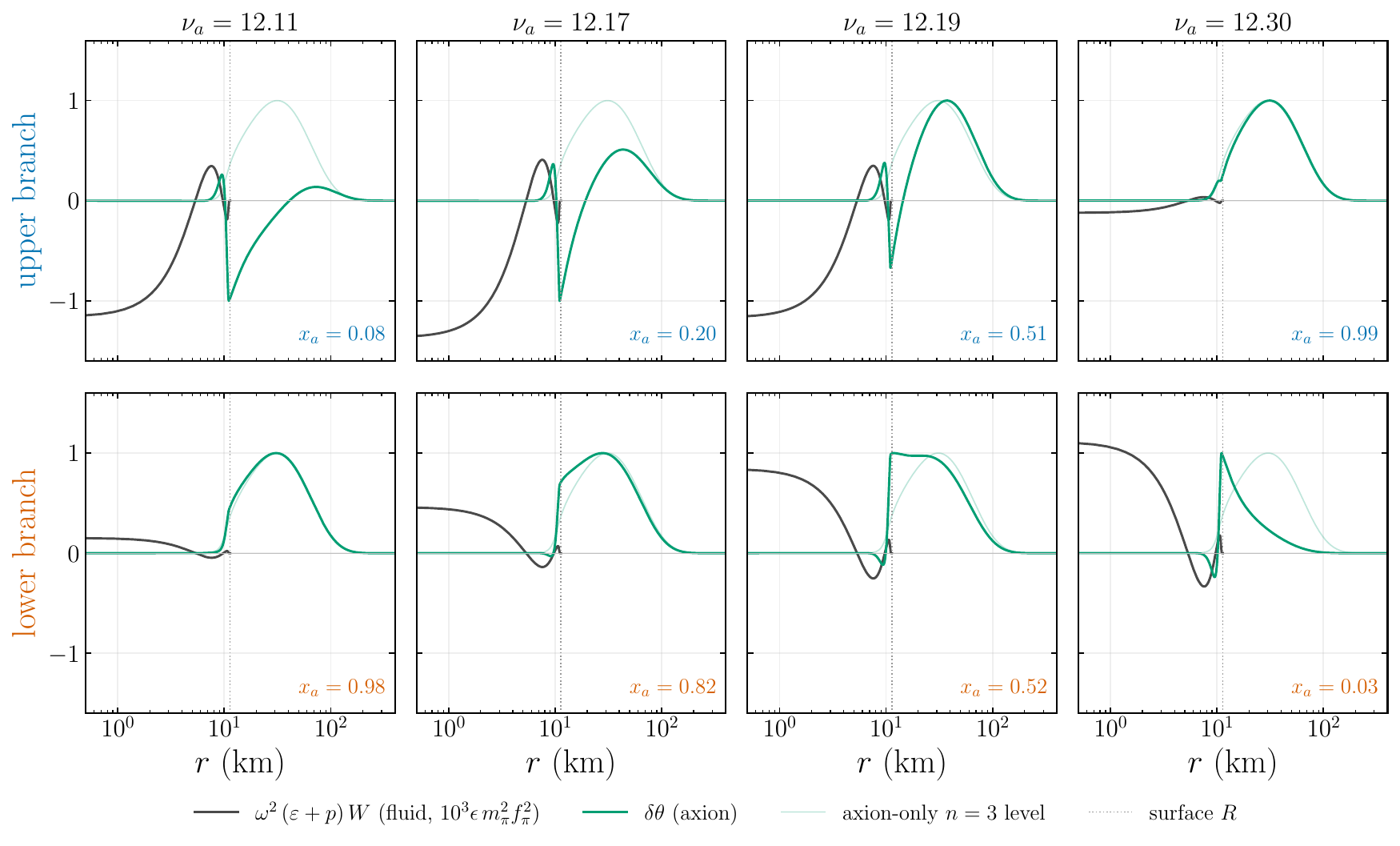}
    \caption{Eigenfunctions of the two coupled branches of
Fig.~\ref{fig:p2_n3_crossing} at four axion masses across the avoided crossing: the
fluid variable $\omega^2(\varepsilon+p)W$ in units of $10^3\,\epsilon\,m_{\pi}^2f_{\pi}^2\simeq 2\times10^3~\rm MeV/fm^3$
(grey) and the axion perturbation $\delta\theta$ (green), each eigenfunction
normalised to $\max|\delta\theta| = 1$. The thin green line is the $n = 3$ level of
the axion-only problem and the dotted vertical line the stellar surface. The axion
energy fraction $x_a$ of each state is given in the lower right corner. Along the upper
branch the state goes from the $p_2$ to the $n = 3$ orbital, along the lower
one the reverse. The narrow feature of $\delta\theta$ at the surface on the
fluid-like states is the field sourced by the fluid through the axion-matter coupling term $\delta n_b\partial m_N^*/\partial\theta
\neq 0$.}

    \label{fig:p2_n3_crossing_eigenfunctions}
\end{figure}

Figure~\ref{fig:p2_n3_crossing_eigenfunctions} shows the eigenfunctions of
both branches at four masses across the avoided crossing. 
Far from it, each state is
either the $p_2$, with the axion perturbation confined to a thin layer at the
surface, or the $n = 3$ orbital at $r \sim 30$~km with a negligible fluid
displacement. At $\nu_a = 12.19$~kHz both carry the fluid mode and the orbital
with comparable weight and opposite relative sign. The surface layer of
$\delta\theta$ visible on the fluid-like states is the axion field sourced by the
fluid's density perturbation in the shell where $\partial m_N^*/\partial\theta
\neq 0$ [see Eq.~\eqref{eq:nonrad_KG_int_main}].

\section{Conclusions}
\label{sec:conclusions}

The formation of a light QCD axion condensate in \acp{ns} can affect not only their equilibrium structure, but also their dynamical properties. Building upon our previous study
of radial oscillations \cite{gomez-banon2026}, we have 
investigated here the impact of the axion condensate on the nonradial oscillation spectrum. 
These modes are particularly relevant for \ac{gw} asteroseismology in \acp{ns}, since their frequencies and damping times encode information on both the stellar structure and the additional degrees of freedom coupled to the fluid.

We have extracted the quadrupole $(l=2)$ oscillation spectrum of a relativistic barotropic \ac{ns} with an axion condensate, using the BSk26 \ac{eos}. The fluid perturbation equations are supplemented by the \ac{kg} equation, dictating the propagation of the axion perturbations. After imposing the appropriate boundary conditions, we solved the resulting eigenvalue problem and identified the complex eigenfrequencies of the modes. For simplicity, we have employed the Cowling approximation, in which spacetime perturbations are neglected, thus enabling the isolation of the damping due solely to axion radiation. The mode spectrum has subsequently been obtained while varying the axion mass within a range relevant for the considered mode frequencies.

As in the case of radial modes, we observed both a modification of the nonradial spectrum due to the presence of the axion field, as well as a new family of axion-dominated modes. The ordinary fluid modes, namely the (fundamental) $f$-modes and the (pressure) $p$-modes, are shifted due to the mere presence of the axion in the background, but also due to their coupling with the propagating axion field. The former leads to an overall increase in the mode frequencies, which is partially compensated for by the latter. The most important consequence, however, is the emergence of a novel damping channel, due to the emission of axion radiation, when the mode frequencies lie above the threshold frequency set by the axion mass, $\mathrm{Re}\,\nu>\nu_a$.

With a frequency of order a few kHz, the $f$-mode never surpasses this threshold, as opposed to the $p$-modes, which radiate axion perturbations efficiently over most of the range, although they may become practically undamped for certain values of $\nu_a$. This makes the condition $\mathrm{Re}\,\nu>\nu_a$ necessary for axion radiation to occur, but not sufficient, as it does not by itself guarantee the transfer of energy from the mode to the axion field. Interestingly, the derived $p$-mode damping times due to axion radiation can become substantially shorter than their \ac{gw} counterparts, for a wide range of values of $\nu_a$.

The coupled system also admits a distinct set of axion-dominated \acp{qnm}. These modes are characterized by much larger imaginary parts than the fluid modes, typically of order kHz, and are therefore strongly damped, showing similarities to the spacetime $w$-modes. The frequencies and the damping rates of the first two axion-dominated modes are of order a few to tens of kHz. In addition, we find an additional class of non-radiative modes, corresponding to gravitationally bound states 
of the scalar field around the star and commonly referred to as gravitational atom levels. It was observed that these modes undergo avoided crossings with $p$-modes, as manifested by a characteristic interchange of axion-led and fluid-led character.

In summary, the present results show that an axion condensate can leave qualitatively distinct signatures in the nonradial oscillation spectrum: structural frequency shifts, dynamical corrections, additional radiative damping, and new axion-dominated modes. Extending the analysis beyond the Cowling approximation, treating the fluid as non-barotropic, and studying more systematically the interaction between the fluid modes and the gravitationally bound axion states are natural next steps.

\begin{acknowledgments}
    A.G.B. is supported by a CIACIF 2022 fellowship, cofunded by the Generalitat Valenciana and the European Union through the European Social Fund. We acknowledge financial support from the Generalitat Valenciana
through PROMETEO PROJECT
CIPROM/2022/13 and from MCIN/AEI (10.13039/501100011033) through Project
PID2025-171322NB-C21.
\end{acknowledgments}

\begin{appendix}

\section{Exterior axion integration and the surface boundary condition}
\label{app:exterior}

In this appendix we describe the construction of the exterior axion solution
used to impose the surface condition~\eqref{eq:theta_surf}. 
The numerical procedure follows the phase-amplitude method introduced for radial
oscillations in Paper~I, with two modifications relevant to the present
problem: the inclusion of the angular contribution for $l\neq0$, and an
improved asymptotic initialization near the propagation threshold $k=0$.
Only these extensions are discussed in detail below; we refer to Paper~I for
the general formulation and numerical motivation of the method.

Throughout this appendix we use the dimensionless variables
\begin{equation}\label{eq:tildevars}
    \tilde r = m_a r,\qquad
    \tilde k = k/m_a.
\end{equation}

We impose the outgoing boundary condition on the axion perturbation by
integrating Eq.~\eqref{eq:nonrad_KG_ext_main} inward from a large radius
$\tilde r_{\infty}$ using the phase--amplitude method, as in Paper~I. Appendix~\ref{app:exterior_pa} summarizes the phase--amplitude formulation
and Appendix~\ref{app:exterior_ic} the two prescriptions for its initial
conditions.

\subsection{Phase-amplitude integration of the exterior solution}
\label{app:exterior_pa}
At sufficiently large radii, the axion background is exponentially
suppressed and its contribution to the exterior \ac{kg}
equation~\eqref{eq:nonrad_KG_ext_main} can be neglected. Applying the change of
variables
\begin{equation}\label{eq:liouville}
    \delta \theta_{\rm ext}=F^{-1/2}\frac{\Psi}{\tilde r},
    \qquad
    F(\tilde r)=1-\frac{2\alpha}{\tilde r},
\end{equation}
removes the first-derivative term,
yielding
\begin{equation}\label{eq:Psieq}
    \frac{d^2\Psi}{d\tilde r^2}
    +\mathcal{U}(\tilde r)\Psi=0,
\end{equation}
where
\begin{equation}\label{eq:Upot}
    \mathcal{U}(\tilde r)
    =\frac{1}{F^2}\left[
        \tilde k^2
        +\frac{2\alpha}{\tilde r}
        -\frac{l(l+1)}{\tilde r^2}F
        +\left(\frac{\alpha}{\tilde r^2}\right)^2
    \right],
\end{equation}
and $\alpha$ is defined in Eq.~\eqref{eq:bohr}. The only modification with respect to the radial effective potential of
Paper~I is the angular contribution
$-l(l+1)F/\tilde r^2$.

As in Paper~I, we follow Andersson, Kokkotas, and Schutz~\cite{andersson1995}
and represent the two independent solutions of Eq.~\eqref{eq:Psieq} as
\begin{equation}\label{eq:PsiQ}
    \Psi_{\pm}=Q^{-1/2}\exp\left(\pm i\,w\right),
    \qquad
    w=\int^{\tilde r}Q(\tilde r')\,d\tilde r',
\end{equation}
so that $\Psi_+\Psi_-=1/Q$. Substitution into
Eq.~\eqref{eq:Psieq} gives
\begin{equation}\label{eq:milne}
    \frac{d^2Q}{d\tilde r^2}
    =
    \frac{3}{2Q}
    \left(\frac{dQ}{d\tilde r}\right)^2
    +2\mathcal{U}Q-2Q^3.
\end{equation}
From Eq.~\eqref{eq:PsiQ}, the logarithmic derivatives of the two independent solutions are
\begin{equation}\label{eq:Lpm}
    \mathcal{L}_{\pm}
    \equiv\frac{1}{\Psi_{\pm}}\frac{d\Psi_{\pm}}{d\tilde r}
    =
    \pm iQ-\frac{1}{2Q}\frac{dQ}{d\tilde r}.
\end{equation}
The initial conditions for $Q$ and $dQ/d\tilde r$ are chosen so that
$\Psi_+$ ($\Psi_-$) corresponds asymptotically to the outgoing (ingoing)
solution, as described in Appendix~\ref{app:exterior_ic}.

\begin{figure}[!htb]
\centering
\begin{tikzpicture}[>=stealth]
  \draw[->] (-0.4,0) -- (9,0) node[below] {$\mathrm{Re}\left(\tilde r\right)$};
  \draw[->] (0,-0.6) -- (0,3.2) node[left] {$\mathrm{Im}\left(\tilde r\right)$};
  \node[below left] at (0,-0.05) {$0$};
  \filldraw (2.6,0) circle (1.6pt);
  \node[below] at (2.6,-0.1) {$\tilde R$};
  \filldraw (4,0) circle (1.6pt);
  \node[below] at (4,-0.12) {$\tilde r_m$};
  \draw[dashed] (4,0) -- (8.8,0);
  \draw[very thick,<-] (4,0) -- (8.6,2.3);
    \node[rotate=26.57, anchor=west] at (8.6,2.3) {$\rho\to\infty$}; 
  \draw (5.1,0) arc (0:26.6:1.1);
  \node at (5.75,0.3) {$\varphi_0$};
\end{tikzpicture}
\caption{Integration contour~\eqref{eq:contour} in the complex $\tilde r$
plane for $\mathrm{Im}\,\tilde k<0$. The path reaches the real $\tilde r$
axis at the matching point $\tilde r_{\rm m}$ along the ray
$\tan\varphi_0=-\mathrm{Im}\,\tilde k/\mathrm{Re}\,\tilde k$, for which the asymptotic solutions of Eq.~\eqref{eq:Psieq} are purely
oscillatory. For $\mathrm{Im}\,\tilde k\geq0$, the integration is instead
performed along the real $\tilde r$ axis ($\varphi_0=0$).}
\label{fig:pacontour}
\end{figure}
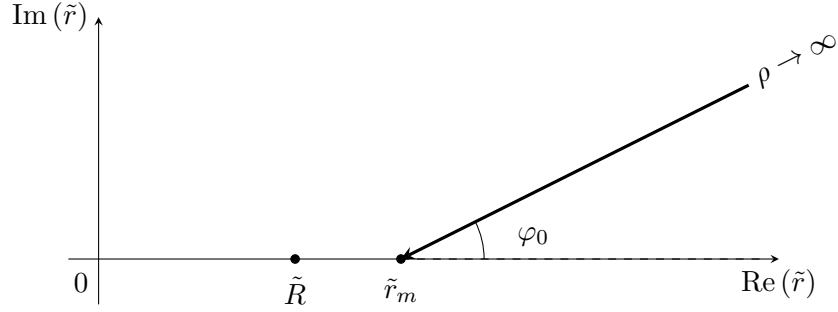

The phase-amplitude integration is terminated at an exterior matching radius
$\tilde r_{\rm m}$, chosen where the axion background is sufficiently
suppressed for Eq.~\eqref{eq:Psieq} to be valid. 
For $\mathrm{Im}\,\tilde k<0$, direct inward integration along the real
axis exponentially amplifies contamination by the unwanted asymptotic
solution. As in Paper~I, we therefore integrate along the complex contour
\begin{equation}
\label{eq:contour}
    \tilde r=\tilde r_{\rm m}+\rho e^{i\varphi_0},
\end{equation}
with
\begin{equation}
    \varphi_0=
    \arctan\left(-\frac{\mathrm{Im}\,\tilde k}
                       {\mathrm{Re}\,\tilde k}\right),
\end{equation}
for which the two asymptotic solutions are oscillatory along the integration
path (see Fig.~\ref{fig:pacontour}). 
For $\mathrm{Im}\,\tilde k\geq0$, no contour rotation is required and
we simply set $\varphi_0=0$.

At $\tilde r_{\rm m}$ the outgoing solution $\Psi_+$ is converted back to the
logarithmic derivative of $\delta\theta_{\rm ext}$ through
Eq.~\eqref{eq:liouville},
\begin{equation}\label{eq:Lext_from_Lplus}
    \mathcal{L}_{\rm ext}
    = m_a\left[\mathcal{L}_+ - \frac{1}{\tilde r} - \frac{\alpha}{\tilde r^2 F}\right],
\end{equation}
which provides the initial value for the integration of the full exterior
equation~\eqref{eq:nonrad_KG_ext_main} from $\tilde r_{\rm m}$ to the
stellar surface along the real radial axis, thereby restoring the
contribution of the nonvanishing axion background. This yields
$\mathcal{L}_{\rm ext}(R)$, the quantity entering the boundary
condition~\eqref{eq:theta_surf}.

\subsection{Asymptotic initialization near the propagation threshold}
\label{app:exterior_ic}
To integrate Eq.~\eqref{eq:milne} inward, initial values for
$Q$ and $dQ/d\tilde r$ must be specified at $\tilde r_\infty$ so that
$\Psi_\pm$ reproduce the asymptotic outgoing and ingoing solutions. In
Paper~I, we used the leading-order WKB approximation,
\begin{equation}\label{eq:lead_wkb}
    Q_{\infty}=\left.\sqrt{\mathcal U}\right|_{\infty},
    \qquad
    \left.\frac{dQ}{d\tilde r}\right|_{\infty}
    =\left.\frac{1}{2\sqrt{\mathcal U}}
    \frac{d\mathcal U}{d\tilde r}\right|_{\infty},
\end{equation}
where the branch of $\sqrt{\mathcal U}$ is chosen to approach the outgoing
wavenumber $+\tilde k$ as $\tilde r\rightarrow\infty$.
This leading-order WKB initialization is adequate away from the propagation
threshold. However, its convergence becomes inefficient as
$\tilde k\rightarrow0$, because the long-range gravitational $1/\tilde r$
term in Eq.~\eqref{eq:Upot} remains comparable to $\tilde k^2$ out to a
characteristic radius
\begin{equation}
    \tilde r_{\rm C}
    \sim
    \frac{2\alpha|2\tilde k^2+1|}{|\tilde k|^2}
    \xrightarrow[\tilde k\to0]{}
    \frac{2\alpha}{|\tilde k|^2},
\end{equation}
so progressively larger values of $\tilde r_\infty$ are required to achieve
the same accuracy with~\eqref{eq:lead_wkb}. To improve the initial conditions
near $\tilde k=0$, we instead use the asymptotic Whittaker solutions of the
massive \ac{kg} equation~\cite{rowan1976,rowan1977}.

Retaining $\mathcal U$ through order $1/\tilde r^2$ gives
\begin{equation}\label{eq:coulomb}
    \frac{d^2\Psi}{d\tilde r^2}
    +\left[
        \tilde k^2
        +\frac{2\alpha(2\tilde k^2+1)}{\tilde r}
        +\frac{4\alpha^2(3\tilde k^2+2)-l(l+1)}{\tilde r^2}
    \right]\Psi=0.
\end{equation}
Defining
\begin{equation}\label{eq:kappanu}
    z=-2i\tilde k\tilde r,
    \qquad
    \kappa=\frac{i\alpha\left(2\tilde k^2+1\right)}{\tilde k},
    \qquad
    \mu=\sqrt{\left(l+\tfrac{1}{2}\right)^2
    -4\alpha^2\left(3\tilde k^2+2\right)},
\end{equation}
reduces Eq.~\eqref{eq:coulomb} to the Whittaker equation. We choose the
initial conditions for $Q$ so that the phase--amplitude solutions $\Psi_\pm$
reproduce asymptotically the outgoing and ingoing Whittaker solutions,
\begin{equation}\label{eq:whitpair}
    \Psi_+\sim W_{\kappa,\mu}(z),
    \qquad
    \Psi_-\sim W_{-\kappa,\mu}(-z).
\end{equation}
Here, $W_{\kappa,\mu}(z)$ denotes the Whittaker function with outgoing asymptotic
behavior
\begin{equation}
    W_{\kappa,\mu}(z)\sim e^{-z/2}z^\kappa .
\end{equation}
The Whittaker approximation incorporates the long-range $1/\tilde r$ tail
explicitly, while its truncated potential differs from the exact potential
$\mathcal U$ only at order $O(\tilde r^{-3})$ and higher.

At $\tilde r_\infty$, we define the logarithmic derivatives of the
corresponding Whittaker solutions as
\begin{equation}
    \mathcal{L}_+^{\rm W}
    =\frac{d}{d\tilde r}\ln W_{\kappa,\mu}(z),
    \qquad
    \mathcal{L}_-^{\rm W}
    =\frac{d}{d\tilde r}\ln W_{-\kappa,\mu}(-z).
\end{equation}
Requiring the logarithmic derivatives of $\Psi_\pm$ in
Eq.~\eqref{eq:Lpm} to match $\mathcal{L}_\pm^{\rm W}$ then determines the initial
values
\begin{equation}\label{eq:whitseed}
    Q_\infty=\frac{\mathcal{L}_+^{\rm W}-\mathcal{L}_-^{\rm W}}{2i},
    \qquad
    \left.\frac{dQ}{d\tilde r}\right|_\infty
    =-Q_\infty\left(\mathcal{L}_+^{\rm W}+\mathcal{L}_-^{\rm W}\right).
\end{equation}
The subsequent phase--amplitude inward integration uses the exact potential
$\mathcal U$ of Eq.~\eqref{eq:Upot}.

In practice, both prescriptions are evaluated at $\tilde r_\infty$. The
accuracy of the WKB approximation is estimated from its residual in
Eq.~\eqref{eq:milne}, while that of the Whittaker approximation is estimated
from the difference between the exact potential~\eqref{eq:Upot} and its
$O(\tilde r^{-2})$ truncation~\eqref{eq:coulomb}. We adopt the prescription with the smaller
local error estimate.

\section{Axion luminosity and mode energy}\label{app:power}
This appendix sketches the derivation of Eqs.~\eqref{eq:La} and~\eqref{eq:La_balance} and
defines the axion energy fraction $x_a$ used in Sec.~\ref{sec:avoid_cross}.

\subsection{Energy balance}
The energy-momentum tensor is the sum of the fluid and axion pieces,
\begin{equation}
    T^{\mu}{}_{\nu} = (\varepsilon+p)\,u^\mu u_\nu - p\,\delta^{\mu}{}_{\nu}
    + f_a^2\,\partial^\mu\theta\,\partial_\nu\theta
    -\delta^{\mu}{}_{\nu}\left[\frac{f_a^2}{2}\,
    \partial^\alpha\theta\,\partial_\alpha\theta-V(\theta)\right],
\end{equation}
where the fluid piece depends on $\theta$ through the effective nucleon mass. 
Consequently, the fluid and axion contributions are not separately conserved, while the total energy-momentum tensor satisfies $\nabla_\mu T^{\mu}{}_{\nu} = 0$.
In the Cowling approximation the metric~\eqref{eq:stat_metric} is unperturbed, so
$\partial_t$ is a Killing vector and the associated energy current $T^{\mu}{}_{t}$ is exactly
conserved, $\partial_\mu(\sqrt{-g}\,T^{\mu}{}_{t})=0$. Integrating over a sphere
of radius $r>R$ gives
\begin{equation}\label{eq:ebalance}
    \frac{dE}{dt}=-L,
\end{equation}
where
\begin{subequations}
\begin{align}
    E(r,t)&=\int\sqrt{-g}\,T^t{}_t\,d^3x,\\
    L(r,t)&=\oint\sqrt{-g}\,T^r{}_t\,d\vartheta\,d\varphi.
\end{align}
\end{subequations}
Thus $L>0$ denotes outward energy flux. Outside the star, only the axion contributes,
\begin{equation}\label{eq:L}
    L(r,t)=-f_a^2r^2e^{\Phi-\lambda}
    \oint\theta'\dot\theta\,d\Omega.
\end{equation}

\subsection{Luminosity and energy of a single mode}
Consider a single mode with complex frequency $\omega$ written in the form
\begin{equation}\label{eq:split}
    \theta=\theta_0(r)+\mathrm{Re}\!\left[
        \delta\theta_{\rm ext}(r)Y_{lm}e^{-i\omega t}
    \right],
\end{equation}
with $\delta\theta_{\rm ext}=(r/R)^l\delta\theta$ extended to the
interior and $\oint|Y_{lm}|^2\,d\Omega=1$.
The fluid perturbations and displacements are defined in
Sec.~\ref{subsec:perturbations}.
Because the background is spherically symmetric, all first-order contributions to the integrated energy and flux vanish for $l\geq1$ after angular integration. The leading mode contributions are therefore quadratic. For
$|\mathrm{Im}\,\omega|\ll\mathrm{Re}\,\omega$, we average over the rapid oscillation phase at a common instant $t_0$, 
neglecting the fractional change in the amplitude over a single oscillation period.

For the flux, the first-order term
$\theta_0'\,\delta\dot\theta$ in Eq.~\eqref{eq:L} integrates to zero over the sphere. 
Averaging the quadratic contribution over the rapid oscillation phase gives, to leading order
\begin{equation}\label{eq:La_app}
    L_a(r) \simeq \frac{f_a^2}{2}\,r^2e^{\Phi-\lambda}\,
    \mathrm{Re}\,\omega\;\mathrm{Im}\,\mathcal{L}_{\rm ext}(r)\,
    |\delta\theta_{\rm ext}|^2 .
\end{equation}

For the energy we use the slow-damping estimate of energy-balance
calculations of \ac{gw} damping times~\cite{thorne1969}: the mode energy is twice the averaged kinetic energy, and for a
weakly damped mode this holds to leading order in
$\mathrm{Im}\,\omega/\mathrm{Re}\,\omega$. These energies, as measured by static observers, redshifted, and integrated over proper volume are
\begin{subequations}\label{eq:mode_energies}
\begin{align}
    E_{\rm fluid} &= \frac{|\omega|^2}{2}\int_0^R (\varepsilon+p)\,e^{\lambda-\Phi}
      \left(\frac{r}{R}\right)^{2l}\left[|W|^2+l(l+1)|\mathcal V|^2\right]dr,\\
    E_a(r) &= \frac{|\omega|^2}{2}\,f_a^2\int_0^{r}e^{\lambda-\Phi}\,
      |\delta\theta_{\rm ext}|^2\,r'^2\,dr',
\end{align}
\end{subequations}
and $E_{\rm mode}(r)\simeq E_{\rm fluid}+E_a(r)$. Both $L_a$ and $E_{\rm mode}$ carry
a factor $e^{2\,\mathrm{Im}\,\omega\,t_0}$, omitted here.

\subsection{Damping rate and axion energy fraction}
Under these assumptions, averaging Eq.~\eqref{eq:ebalance} gives
\begin{equation}\label{eq:Imomega_app}
    \mathrm{Im}\,\omega\simeq-\frac{L_a(r)}{2E_{\rm mode}(r)}.
\end{equation}
Both $L_a(r)$ and $E_{\rm mode}(r)$ contain the 
common factor $e^{2\,\mathrm{Im}\,\omega\,t_0}$ 
which cancels in their ratio.  
In the asymptotic wave zone, the outgoing solution has the leading radial dependence
$\delta\theta_{\rm ext}\simeq A_{\rm out}\,e^{ikr}/r$. Consequently,
$\mathrm{Im}\,\mathcal{L}_{\rm ext}\simeq\mathrm{Re}\,k$ 
to leading order and
$r^2|\delta\theta_{\rm ext}|^2\simeq|A_{\rm out}|^2e^{-2\,\mathrm{Im}\,k\,r}$,
which gives Eq.~\eqref{eq:La}. 

Finally, for the bound modes, for which $\mathrm{Im}\,\omega=0$ and the axion profile is normalizable, we define the axion-content diagnostic
\begin{equation}\label{eq:xa}
    x_a=\frac{E_a}{E_a+E_{\rm fluid}},
\end{equation}
where $E_a$ is evaluated over the full axion perturbation profile. $x_a$ provides a convenient measure of their fluid or axion character: $x_a \ll 1$ corresponds to a fluid-dominated mode and $x_a \lesssim 1$ to an axion-dominated state.

\end{appendix}

\bibliographystyle{apsrev4-2}
\bibliography{common/Thesis}

\end{document}